\documentclass[12pt]{article}
 
\usepackage{amsmath,amssymb}
\usepackage{bm}
\usepackage{graphicx, color}
\usepackage{wrapfig}
\usepackage{cite}
\begin{document}
\title{
\begin{flushright}
\ \\*[-80pt] 
\begin{minipage}{0.22\linewidth}
\normalsize
APCTP Pre2020-008
 \\*[50pt]
\end{minipage}
\end{flushright}
{\Large \bf 
	 Quark and lepton flavors with  common modulus $\tau$\\  
 in $A_4$ modular symmetry 	\\*[20pt]}}

\author{ 
\centerline{
Hiroshi Okada $^{a,b}\footnote{E-mail address: hiroshi.okada@apctp.org}$~ and 
~~Morimitsu Tanimoto $^{c}\footnote{E-mail address: tanimoto@muse.sc.niigata-u.ac.jp}$} \\*[5pt]
\centerline{
\begin{minipage}{\linewidth}
\begin{center}
$^a${\it \normalsize
Asia Pacific Center for Theoretical Physics, Pohang 37673, Republic of Korea} \\*[5pt]
$^b${\it \normalsize
Department of Physics, Pohang University of Science and Technology, Pohang 37673,\\Republic of Korea} \\*[5pt]
$^c${\it \normalsize
Department of Physics, Niigata University, Niigata 950-2181, Japan}
\end{center}
\end{minipage}}
\\*[50pt]}

\date{
\centerline{\small \bf Abstract}
\begin{minipage}{0.9\linewidth}
\medskip 
\medskip 
\small 
We study quark and lepton mass matrices 
with the common  modulus $\tau$ in  the  $A_4$ modular symmetry.
The viable quark mass matrices are composed 
of modular forms of weights $2$, $4$ and $6$.
It is remarked that the modulus $\tau$ is close to $i$,
which is a fixed point in the fundamental region of SL$(2,Z)$,
and  the CP symmetry is not violated.
Indeed,   the observed CP violation is reproduced
at $\tau$ which is deviated a little bit from  $\tau=i$.
The  charged lepton mass matrix
is also given by using modular forms of weights $2$, $4$ and $6$,
where five cases   have been examined. 
The neutrino mass matrix is generated  in terms of 
the modular forms of weight $4$  through the Weinberg operator.  
Lepton mass matrices are also consistent with the observed
mixing angles at  $\tau$ close to $i$ for NH of neutrino masses.
Allowed regions of $\tau$ of quarks and  leptons  overlap each other
for all cases of the charged lepton mass matrix.
However, the sum of neutrino masses is  crucial to test
the common $\tau$ for quarks and leptons.
The minimal sum of neutrino masses $\sum m_i$
is  $140$meV at the common $\tau$.
 The inverted  hierarchy of  neutrino masses
is unfavorable in our framework.
It is emphasized that  our result suggests  the  residual symmetry  $\mathbb{Z}_2^{S}=\{ I, S \}$ in the quark and lepton mass matrices.
\end{minipage}
}

\begin{titlepage}
\maketitle
\thispagestyle{empty}
\end{titlepage}

\section{Introduction}

The origin of the flavors  is one of important issues in particle physics.
A lot of  works have been presented by using the  discrete  groups
 for flavors to understand the flavor structures of quarks and leptons.
 In  the early models of quark masses and mixing angles, 
the  $S_3$ symmetry was used 
 \cite{Pakvasa:1977in,Wilczek:1977uh}.
 It was also discussed  to understand the large mixing angle
 \cite{Fukugita:1998vn} in the oscillation of atmospheric neutrinos \cite{Fukuda:1998mi}. 
 For the last twenty years, the  discrete symmetries of flavors have been developed, that is
  motivated by the precise observation of  flavor mixing angles of  leptons
  \cite{Altarelli:2010gt,Ishimori:2010au,Ishimori:2012zz,Hernandez:2012ra,King:2013eh,King:2014nza,Tanimoto:2015nfa,King:2017guk,Petcov:2017ggy,Feruglio:2019ktm}.

Many models have been proposed by using 
the non-Abelian discrete groups  $S_3$, $A_4$, $S_4$, $A_5$ and other groups with larger orders to explain the large neutrino mixing angles.
Among them, the $A_4$ flavor model is attractive one 
because the $A_4$ group is the minimal one including a triplet 
 irreducible representation, 
which allows for a natural explanation of the  
existence of  three families of leptons 
\cite{Ma:2001dn,Babu:2002dz,Altarelli:2005yp,Altarelli:2005yx,
Shimizu:2011xg,Petcov:2018snn,Kang:2018txu}.
However, variety of models is so wide that it is difficult to show 
a clear evidence of the $A_4$ flavor symmetry.

Recently, a new  approach to the lepton flavor problem appeared
based on the invariance of the modular group \cite{Feruglio:2017spp}, 
where the model of the finite
modular group  $\Gamma_3 \simeq A_4$ has been presented.
This work  inspired further studies of the modular invariance 
 to the lepton flavor problem. 
The modular group includes the finite groups $S_3$, $A_4$, $S_4$, and $A_5$ \cite{deAdelhartToorop:2011re}.
Therefore, an interesting framework for the construction of flavor models
has been put forward based on  the $\Gamma_3 \simeq A_4$ modular group \cite{Feruglio:2017spp}, and further, based on $\Gamma_2 \simeq S_3$ \cite{Kobayashi:2018vbk}.
The  flavor models have been proposed by using modular symmetries  
$\Gamma_4 \simeq S_4$ \cite{Penedo:2018nmg} 
and  $\Gamma_5 \simeq A_5$ \cite{Novichkov:2018nkm}. 
Phenomenological discussions of the neutrino flavor mixing have been done
based on  $A_4$ \cite{Criado:2018thu,Kobayashi:2018scp,Ding:2019zxk}, $S_4$ \cite{Novichkov:2018ovf,Kobayashi:2019mna,Wang:2019ovr}, 
$A_5$ \cite{Ding:2019xna}, 
and  $T'$~\cite{Liu:2019khw,Chen:2020udk} modular groups, respectively.
In particular, the comprehensive analysis of the $A_4$ modular group 
has provided a distinct prediction of the neutrino mixing angles and the CP violating phase \cite{Kobayashi:2018scp}.

The $A_4$ modular symmetry has been also applied to
the leptogenesis \cite{Asaka:2019vev}, on the othr hand,
it is discussed  in the SU$(5)$ grand
unified theory (GUT) of  quarks and leptons  \cite{deAnda:2018ecu,Kobayashi:2019rzp}.
The residual symmetry of the $A_4$ modular symmetry has presented the
interesting phenomenology \cite{Novichkov:2018yse}.
Furthermore, modular forms for $\Delta(96)$ and $\Delta(384)$ were constructed \cite{Kobayashi:2018bff},
and the extension of the traditional flavor group  is discussed with modular symmetries \cite{Baur:2019kwi}.
The level $7$ finite modular group $\Gamma_7\simeq PSL(2,Z_7)$
 is also presented for the lepton mixing \cite{Ding:2020msi}.
Moreover, multiple modular symmetries are proposed as the origin of flavor\cite{deMedeirosVarzielas:2019cyj}.
The modular invariance has been also studied combining with the generalized CP symmetries for theories of flavors \cite{Novichkov:2019sqv}.
 The quark mass matrix  has been discussed in the $S_3$ and $A_4$ modular symmetries as well \cite{Kobayashi:2018wkl,Okada:2018yrn,Okada:2019uoy}.
 Besides mass matrices of quarks and leptons,
 related topics have been discussed 
  in the baryon number violation  \cite{Kobayashi:2018wkl}, 
   the dark matter \cite{Nomura:2019jxj, Okada:2019xqk}
   and the modular symmetry anomaly  \cite{Kariyazono:2019ehj}.
  Furthere phenomenology has been developed  in many works
   \cite{Nomura:2019yft,Okada:2019lzv,Nomura:2019lnr,Criado:2019tzk,Gui-JunDing:2019wap,Zhang:2019ngf,Nomura:2019xsb,Kobayashi:2019gtp,Lu:2019vgm,Wang:2019xbo,King:2020qaj,Abbas:2020qzc,Okada:2020oxh,Okada:2020dmb,Ding:2020yen}
    while theoretical investigations are also proceeded \cite{Kobayashi:2019xvz,Nilles:2020kgo}.
 
 
In this work, we study both quarks and leptons
 in the $A_4$ modular symmetry.
 If the flavor of quarks and leptons is originated from a same two-dimensional compact space, 
  quarks and leptons have same flavor symmetry and the common  modulus $\tau$.
  Therefore, it is challenging to reproduce
observed hierarchical three Cabibbo-Kobayashi-Maskawa (CKM) mixing angles and the CP violating phase 
while observed large mixing angles are also reproduced in the lepton sector
within the framework of the $A_4$ modular invariance with the common $\tau$.
This work provides a new aspect for the unification theory
of the quark and lepton flavors.
We have already discussed the quark mass matrices in the $A_4$
 modular symmetry \cite{Kobayashi:2018wkl,Okada:2018yrn,Okada:2019uoy},
 where modular forms of weight $6$ play an important role.
 In this paper, we present the comprehensive analysis by
   adopting  modular forms of weight $4$ and $6$
 in addition to  modular forms of weight $2$ for quarks and charged leptons.
  We  take  modular forms of weight $4$ for the neutrino mass matrix generated by the Weinberg operator.
 We obtain the  successful CKM mixing matrix at $\tau$ close to the fixed point $i$.  We also discuss Pontecorvo-Maki-Nakagawa-Sakata (PMNS) mixing 
 \cite{Maki:1962mu,Pontecorvo:1967fh} 
  around  $\tau=i$ as well as 
  the CP violating Dirac phase  of  leptons
 which is expected to be observed at T2K and NO$\nu$A experiments \cite{T2K:2020,Adamson:2017gxd},
  with reference to the sum of neutrino masses.
  It is found that the sum of neutrino masses is
    crucial  to realize the common $\tau$ for quarks and leptons.


The paper is organized as follows.
In section 2,  we give a brief review on the modular symmetry and 
modular forms of weights $2$, $4$ and $6$. 
In section 3, we present the model for quark mass matrices in the $A_4$
modular symmetry.
In section 4, the modulus $\tau$ is fixed by the CKM matrix.
In section 5, we discuss the lepton mass matrices,
and in section 6, we examine  $\tau$ in the lepton mixing
 and some predictions.
Section 7 is devoted to a summary and discussions.
In Appendix A, the tensor product  of the $A_4$ group is presented.
In Appendix B,  we present how to obtain  Dirac $CP$ phase, Majorana phases and  the effective mass of the $0\nu\beta\beta$ decay.

\section{Modular group and modular forms}

The modular group $\bar\Gamma$ is the group of linear fractional transformation
$\gamma$ acting on the modulus  $\tau$, 
belonging to the upper-half complex plane as:
\begin{equation}\label{eq:tau-SL2Z}
\tau \longrightarrow \gamma\tau= \frac{a\tau + b}{c \tau + d}\ ,~~
{\rm where}~~ a,b,c,d \in \mathbb{Z}~~ {\rm and }~~ ad-bc=1, 
~~ {\rm Im} [\tau]>0 ~ ,
\end{equation}
 which is isomorphic to  $PSL(2,\mathbb{Z})=SL(2,\mathbb{Z})/\{I,-I\}$ transformation.
This modular transformation is generated by $S$ and $T$, 
\begin{eqnarray}
S:\tau \longrightarrow -\frac{1}{\tau}\ , \qquad\qquad
T:\tau \longrightarrow \tau + 1\ ,
\label{symmetry}
\end{eqnarray}
which satisfy the following algebraic relations, 
\begin{equation}
S^2 =\mathbb{I}\ , \qquad (ST)^3 =\mathbb{I}\ .
\end{equation}

 We introduce the series of groups $\Gamma(N)~ (N=1,2,3,\dots)$,
   called principal congruence subgroups, defined by
 \begin{align}
 \begin{aligned}
 \Gamma(N)= \left \{ 
 \begin{pmatrix}
 a & b  \\
 c & d  
 \end{pmatrix} \in SL(2,\mathbb{Z})~ ,
 ~~
 \begin{pmatrix}
  a & b  \\
 c & d  
 \end{pmatrix} =
  \begin{pmatrix}
  1 & 0  \\
  0 & 1  
  \end{pmatrix} ~~({\rm mod} N) \right \}
 \end{aligned}\, .
 \end{align}
 For $N=2$, we define $\bar\Gamma(2)\equiv \Gamma(2)/\{I,-I\}$.
Since the element $-I$ does not belong to $\Gamma(N)$
  for $N>2$, we have $\bar\Gamma(N)= \Gamma(N)$.
  The quotient groups defined as
  $\Gamma_N\equiv \bar \Gamma/\bar \Gamma(N)$
  are  finite modular groups.
In this finite groups $\Gamma_N$, $T^N=\mathbb{I}$  is imposed.
 The  groups $\Gamma_N$ with $N=2,3,4,5$ are isomorphic to
$S_3$, $A_4$, $S_4$ and $A_5$, respectively \cite{deAdelhartToorop:2011re}.

Modular forms of  level $N$ are 
holomorphic functions $f(\tau)$  transforming under the action of $\Gamma(N)$ as:
\begin{equation}
f(\gamma\tau)= (c\tau+d)^kf(\tau)~, ~~ \gamma \in \Gamma(N)~ ,
\end{equation}
where $k$ is the so-called as the  modular weight.

Superstring theory on the torus $T^2$ or orbifold $T^2/Z_N$ has the modular symmetry \cite{Lauer:1989ax,Lerche:1989cs,Ferrara:1989qb,Cremades:2004wa,Kobayashi:2017dyu,Kobayashi:2018rad}.
Its low energy effective field theory is described in terms of  supergravity theory,
and  string-derived supergravity theory has also the modular symmetry.
Under the modular transformation of Eq.(\ref{eq:tau-SL2Z}), chiral superfields $\phi^{(I)}$ 
transform as \cite{Ferrara:1989bc},
\begin{equation}
\phi^{(I)}\to(c\tau+d)^{-k_I}\rho^{(I)}(\gamma)\phi^{(I)},
\end{equation}
where  $-k_I$ is the modular weight and $\rho^{(I)}(\gamma)$ denotes an unitary representation matrix of $\gamma\in\Gamma(N)$.

 In the present article we study global supersymmetric models, e.g., 
minimal supersymmetric extensions of the Standard Model (MSSM).
The superpotential which is built from matter fields and modular forms
is assumed to be modular invariant, i.e., to have 
a vanishing modular weight. For given modular forms 
this can be achieved by assigning appropriate
weights to the matter superfields.

The kinetic terms  are  derived from a K\"ahler potential.
The K\"ahler potential of chiral matter fields $\phi^{(I)}$ with the modular weight $-k_I$ is given simply  by 
\begin{equation}
K^{\rm matter} = \frac{1}{[i(\bar\tau - \tau)]^{k_I}} |\phi^{(I)}|^2,
\end{equation}
where the superfield and its scalar component are denoted by the same letter, and  $\bar\tau =\tau^*$ after taking the vacuum expectation value (VEV).
Therefore, 
the canonical form of the kinetic terms  is obtained by the overall normalization of the quark and lepton mass matrices
\footnote{The most general K\"ahler potential consistent with the modular symmetry possibly contains additional terms, as
	recently pointed out in Ref. \cite{Chen:2019ewa}. However, we consider only the simplest form of
	the K\"ahler potential.}.

For $\Gamma_3\simeq A_4$, the dimension of the linear space 
${\cal M}_k(\Gamma_3)$ 
of modular forms of weight $k$ is $k+1$ \cite{Gunning:1962,Schoeneberg:1974,Koblitz:1984}, i.e., there are three linearly 
independent modular forms of the lowest non-trivial weight $2$.
These forms have been explicitly obtained \cite{Feruglio:2017spp} in terms of
the Dedekind eta-function $\eta(\tau)$: 
\begin{equation}
\eta(\tau) = q^{1/24} \prod_{n =1}^\infty (1-q^n)~, 
 \quad\qquad  q= \exp \ (i 2 \pi  \tau )~,
\end{equation}
%
where $\eta(\tau)$ is a  so called  modular form of weight~$1/2$. 
In what follows we will use the following base of the 
$A_4$ generators  $S$ and $T$ in the triplet representation:
\begin{align}
\begin{aligned}
S=\frac{1}{3}
\begin{pmatrix}
-1 & 2 & 2 \\
2 &-1 & 2 \\
2 & 2 &-1
\end{pmatrix},
\end{aligned}
\qquad \qquad
\begin{aligned}
T=
\begin{pmatrix}
1 & 0& 0 \\
0 &\omega& 0 \\
0 & 0 & \omega^2
\end{pmatrix}, 
\end{aligned}
\label{STbase}
\end{align}
%
where $\omega=\exp (i\frac{2}{3}\pi)$ .
The  modular forms of weight 2 transforming
as a triplet of $A_4$ can be written in terms of 
$\eta(\tau)$ and its derivative \cite{Feruglio:2017spp}:
\begin{eqnarray} 
\label{eq:Y-A4}
Y_1(\tau) &=& \frac{i}{2\pi}\left( \frac{\eta'(\tau/3)}{\eta(\tau/3)}  +\frac{\eta'((\tau +1)/3)}{\eta((\tau+1)/3)}  
+\frac{\eta'((\tau +2)/3)}{\eta((\tau+2)/3)} - \frac{27\eta'(3\tau)}{\eta(3\tau)}  \right), \nonumber \\
Y_2(\tau) &=& \frac{-i}{\pi}\left( \frac{\eta'(\tau/3)}{\eta(\tau/3)}  +\omega^2\frac{\eta'((\tau +1)/3)}{\eta((\tau+1)/3)}  
+\omega \frac{\eta'((\tau +2)/3)}{\eta((\tau+2)/3)}  \right) , \label{Yi} \\ 
Y_3(\tau) &=& \frac{-i}{\pi}\left( \frac{\eta'(\tau/3)}{\eta(\tau/3)}  +\omega\frac{\eta'((\tau +1)/3)}{\eta((\tau+1)/3)}  
+\omega^2 \frac{\eta'((\tau +2)/3)}{\eta((\tau+2)/3)}  \right)\,,
\nonumber
\end{eqnarray}
%
which have the following  $q$-expansions:
\begin{align}
{\bf Y^{(2)}_3}
=\begin{pmatrix}Y_1(\tau)\\Y_2(\tau)\\Y_3(\tau)\end{pmatrix}=
\begin{pmatrix}
1+12q+36q^2+12q^3+\dots \\
-6q^{1/3}(1+7q+8q^2+\dots) \\
-18q^{2/3}(1+2q+5q^2+\dots)\end{pmatrix}.
\label{Y(2)}
\end{align}
%
They satisfy also the constraint \cite{Feruglio:2017spp}:
\begin{align}
(Y_2(\tau))^2+2Y_1(\tau) Y_3(\tau)=0~.
\label{condition}
\end{align}

The  modular forms of the  higher weight, $k$, can be obtained
by the $A_4$ tensor products of  the modular forms  
 ${\bf Y^{(2)}_3}$,   as given in Appendix A.
  For weight $4$, that is $k=4$, there are  five modular forms
   by the tensor product of  $\bf 3\otimes 3$ as:
\begin{align}
&\begin{aligned}
{\bf Y^{(4)}_1}=Y_1^2+2 Y_2 Y_3 \, , \quad
{\bf Y^{(4)}_{1'}}=Y_3^2+2 Y_1 Y_2 \, , \quad
{\bf Y^{(4)}_{1''}}=Y_2^2+2 Y_1 Y_3=0 \, , \quad
\end{aligned}\nonumber \\
\nonumber \\
&\begin{aligned} {\bf Y^{(4)}_{3}}=
\begin{pmatrix}
Y_1^{(4)}  \\
Y_2^{(4)} \\
Y_3^{(4)}
\end{pmatrix}
=
\begin{pmatrix}
Y_1^2-Y_2 Y_3  \\
Y_3^2 -Y_1 Y_2 \\
Y_2^2-Y_1 Y_3
\end{pmatrix}\, , 
\end{aligned}
\label{weight4}
\end{align}
where ${\bf Y^{(4)}_{1''}}$ vanishes due to the constraint of
 Eq.\,(\ref{condition}).
 For wight 6,  there are  seven modular forms
by the tensor products of  $A_4$ as:
\begin{align}
&\begin{aligned}
{\bf Y^{(6)}_1}=Y_1^3+ Y_2^3+Y_3^3 -3Y_1 Y_2 Y_3  \, , 
\end{aligned} \nonumber \\
\nonumber \\
&\begin{aligned} {\bf Y^{(6)}_3}\equiv 
\begin{pmatrix}
Y_1^{(6)}  \\
Y_2^{(6)} \\
Y_3^{(6)}
\end{pmatrix}
=
\begin{pmatrix}
Y_1^3+2 Y_1 Y_2 Y_3   \\
Y_1^2 Y_2+2 Y_2^2 Y_3 \\
Y_1^2Y_3+2Y_3^2Y_2
\end{pmatrix}\, , \qquad
\end{aligned}
\begin{aligned} {\bf Y^{(6)}_{3'}}\equiv
\begin{pmatrix}
Y_1^{'(6)}  \\
Y_2^{'(6)} \\
Y_3^{'(6)}
\end{pmatrix}
=
\begin{pmatrix}
Y_3^3+2 Y_1 Y_2 Y_3   \\
Y_3^2 Y_1+2 Y_1^2 Y_2 \\
Y_3^2Y_2+2Y_2^2Y_1
\end{pmatrix}\, . 
\end{aligned}
\label{weight6}
\end{align}
 By using these modular forms of weights $2, 4, 6$,
  we discuss  quark and lepton mass matrices.


\section{ $A_4$ modular invariant quark mass matrices}

Let us consider a $A_4$ modular invariant flavor model for quarks.
 There are freedoms for the assignments of irreducible representations and modular weights to quarks and Higgs doublets.
The simplest one is to assign  the triplet of the $A_4$ group 
    to  three left-handed quarks, but three different singlets 
    $\bf (1,1'',1')$ of  $A_4$ to
    the  three right-handed quarks,
    ($u^c, c^c, t^c$) and  ($d^c, s^c, b^c$), respectively,
    where   the sum of weights of the left-handed and the  right-handed quarks is $-2$.

 Then, there appear three independent couplings in the superpotential of 
the up-type  and down-type quark sectors, respectively,
as follows:
 \begin{align}
 w_u&=\alpha_u u^c H_u {\bf Y^{(2)}_3} Q+
 \beta_u c^c H_u {\bf Y^{(2)}_3}Q+
 \gamma_u t^c H_u {\bf Y^{(2)}_3}Q\,,\label{upquark} 
 \end{align}
\begin{align}
w_d&=\alpha_d d^c H_d {\bf Y^{(2)}_3}Q+
\beta_d s^c H_d {\bf Y^{(2)}_3}Q+
\gamma_d b^c H_d {\bf Y^{(2)}_3}Q\,,
\label{downquark}
\end{align}
where $Q$ is the left-handed $A_4$ triplet quarks,
and $H_q$ is the Higgs doublets.
The parameters $\alpha_q$,  $\beta_q$,  $\gamma_q$ ($q=u,d$)
are constant coefficients.
Assign the left-handed $A_4$ triplet $Q$ to $(u_L, c_L, t_L)$ and
$(d_L, s_L, b_L)$.
By using the decomposition of the $A_4$ tensor product in Appendix A, 
the superpotentials in Eqs.(\ref{upquark}) and (\ref{downquark}) give 
the mass matrix of quarks, which is written in terms of
modular forms of weight 2:
\begin{align}
\begin{aligned}
M_q=v_q
\begin{pmatrix}
\alpha_q & 0 & 0 \\
0 &\beta_q & 0\\
0 & 0 &\gamma_q
\end{pmatrix}
\begin{pmatrix}
Y_1 & Y_3& Y_2\\
Y_2 & Y_1 &  Y_3 \\
Y_3&  Y_2&  Y_1
\end{pmatrix}_{RL},     \qquad (q=u, d)~,
\end{aligned}
\label{matrixSM}
\end{align}
where the argument $\tau$ in the modular forms $Y_i(\tau)$ is  omitted.
 The coefficient $v_q$ is the VEV of the Higgs field $H_q$.
Unknown coefficients  $\alpha_q$,  $\beta_q$,  $\gamma_q$  can be adjusted to the  observed quark masses. The remained parameter is only   the modulus, 
$\tau$. 
The numerical study of the quark mass matrix in Eq.(\ref{matrixSM}) is
 rather easy. 
However, it is impossible to  reproduce
observed  hierarchical three CKM mixing angles
by fixing one complex parameter $\tau$.
 
\begin{table}[h]
	\centering
	\begin{tabular}{|c||c|c|c|c|c|c|} \hline
		&$Q$&$(q_{1}^c,q_{2}^c,q_{3}^c)$&$H_q$&
$\bf Y_3^{(6)}, \ Y_{3'}^{(6)}$& $\bf Y_3^{(4)}$& $\bf Y_3^{(2)}$\\  \hline\hline 
		\rule[14pt]{0pt}{0pt}
		$SU(2)$&$\bf 2$&$\bf 1$&$\bf 2$&$\bf 1$&$\bf 1$&$\bf 1$\\
		$A_4$&$\bf 3$& \bf (1,\ 1$''$,\ 1$'$)&$\bf 1$&$\bf 3 $&$\bf 3$&$\bf 3$\\
		$-k_I$&$ -2$&$(-4,\ -2,\ 0)$&0&$k=6$&$k= 4$&$k=2$ \\ \hline
	\end{tabular}
	\caption{Assignments of representations and  weights
		$-k_I$ for MSSM fields and  modular forms.
	}
	\label{tb:weight6}
\end{table}
 In order to obtain  realistic quark mass matrices,
  we use  modular forms of  weight $4$ and  $6$ 
  in addition to   weight $2$  modular forms.
  They are given in Eqs.(\ref{weight4}) and (\ref{weight6}).
   We present the superpotential of the quark sector as follows:
 \begin{align}
w_q&=\alpha_q q_1^c H_q {\bf Y^{(6)}_3} Q+
\alpha'_q q_1^c H_q {\bf Y_{3'}^{(6)}} Q+
\beta_q q_2^c H_q {\bf Y^{(4)}_3}Q+
\gamma_q q_3^c H_q {\bf Y^{(2)}_3}Q\,,
\end{align}
where assignments of representations and  weights
    for MSSM fields are given in  Table 1.
    The quark mass matrix is written as:
\begin{align}
\begin{aligned}
M_q=
\begin{pmatrix}
\alpha_q & 0 & 0 \\
0 &\beta_q & 0\\
0 & 0 &\gamma_q
\end{pmatrix} \left [
\begin{pmatrix}
Y_1^{(6)}+g_q Y_1^{'(6)} & Y_3^{(6)} +g_q Y_3^{'(6)} 
& Y_2^{(6)}+g_q Y_2^{'(6)} \\
Y_2^{(4)} & Y_1^{(4)} &  Y_3^{(4)} \\
Y_3^{(2)} &  Y_2^{(2)}&  Y_1^{(2)}
\end{pmatrix}
\right ]_{RL},
\end{aligned}
\label{matrix6}
\end{align}
where $g_q\equiv \alpha_q'/\alpha_q$\,.
Parameters $\alpha_q$,  $\beta_q$,  $\gamma_q$  are  real,
on the other hand, $g_{q}$ are  complex parameters.
The  parameters of our model are real six parameters,
$\alpha_u$,  $\beta_u$,  $\gamma_u$, $\alpha_d$,  $\beta_d$,  $\gamma_d$,
and  complex parameters $g_u$, $g_d$ in addition to the modulus $\tau$.
Now the model could be reconciled with observed values.
Indeed, we have found parameter sets, which is consistent with
the CKM observables and quark masses, in our numerical results.


\section{Fixing modulus $\tau$ by the CKM mixing}

In order to obtain the left-handed flavor mixing,
we calculate $M_u^{\dagger} M_u$ and $M_d^{\dagger} M_d$, respectively.
At first, we take a random point of  $\tau$ and  $g_q$
which are scanned  in the complex plane 
by generating random numbers. The modulus $\tau$ is scanned
 in the fundamental region of the modular symmetry.
 In practice,  the  scanned range of 
${\rm Im } [\tau]$  is $[\sqrt{3}/2,2]$, in which the lower-cut $\sqrt{3}/2$ is at the cusp of the fundamental region, and  the upper-cut $2$ is enough large for estimating  $Y_i$.
On the other hand,   ${\rm Re } [\tau]$ is scanned in
the fundamental region  $[-1/2, 1/2]$ of the modular group.
We also scan  in 
$|g_u| \in [0,100]$ and $|g_d|\in [0,100]$ while these phases are scanned in $[-\pi,\pi]$.
Then, parameters $\alpha_q$,  $\beta_q$,  $\gamma_q$ ($q=u,d$)
are given in terms of $\tau$ and $g_q$ after inputting six quark masses.

Finally, we calculate three CKM mixing angles
and the CP violating phase in terms of the model parameters $\tau$, $g_u$ and $g_d$.
We keep the parameter sets, in  which the value of each observable is reproduced
within  the three times of $1\sigma$ interval of error-bars.
We continue this procedure to obtain enough points for plotting allowed region.

We input quark masses in order to constrain model parameters.
Since the modulus $\tau$ obtains the expectation value
 by the breaking of the modular invariance at the high mass scale,
   the quark masses are put  at the GUT scale.
   The observed masses and CKM parameters run to the GUT scale
 by the renormalization group equations (RGEs).
In our work, we adopt numerical values  of Yukawa couplings of quarks
  at the GUT scale $2\times 10^{16}$ GeV with $\tan\beta=5$ 
  in the framework of the minimal SUSY breaking scenarios
\cite{Antusch:2013jca, Bjorkeroth:2015ora}:
\begin{align}
\begin{aligned}
&y_d=(4.81\pm 1.06) \times 10^{-6}, \quad y_s=(9.52\pm 1.03) \times 10^{-5}, \quad y_b=(6.95\pm 0.175) \times 10^{-3}, \\
\rule[15pt]{0pt}{1pt}
&y_u=(2.92\pm 1.81) \times 10^{-6}, \quad y_c=(1.43{\pm 0.100}) \times 10^{-3}, \quad y_t=0.534\pm 0.0341  ~~,
\end{aligned}\label{yukawa5}
\end{align}
which give quark masses as $m_q=y_q v_H$ with $v_H=174$ GeV.
In our numerical calculation, we input  $2\sigma$ interval for quark masses.

We also use the following CKM mixing angles to focus on parameter regions consistent with the 
experimental data  at the GUT scale $2\times 10^{16}$ GeV, where $\tan\beta=5$
is taken  \cite{Antusch:2013jca, Bjorkeroth:2015ora}:
 \begin{align}
\begin{aligned}
&\theta_{12}=13.027^\circ\pm 0.0814^\circ ~ , \qquad
\theta_{23}=2.054^\circ\pm 0.384^\circ ~ ,  \qquad
 \theta_{13}=0.1802^\circ\pm 0.0281^\circ~ .
\end{aligned}\label{CKM}
\end{align}
Here $\theta_{ij}$ is given in the PDG notation
of the CKM matrix $V_{\rm CKM}$ \cite{Tanabashi:2018oca}.
The observed CP violating phase is given as:
\begin{equation}
\delta_{CP}=69.21^\circ\pm 6.19^\circ~ ,
\label{CKMphase}
\end{equation}
which is also  in the PDG notation.
The error intervals in Eqs.\,(\ref{yukawa5}),  (\ref{CKM}) and (\ref{CKMphase}) represent $1\sigma$ interval.
 
In our model, we have   three complex parameters,
  $\tau$, $g_u$ and $g_d$
 after inputting six quark masses. 
  The allowed regions of these parameters are obtained by 
  inputting the observed three CKM mixing angles
  and  CP violating phase 
  with three times of $1 \sigma$ interval in Eqs.\,(\ref{CKM}) and (\ref{CKMphase}).
We have succeeded to  reproduce completely four CKM elements 
in the parameter ranges of Table 2.

  
   
   \begin{table}[h!]
   	\centering
   	\begin{tabular}{|c|c|c|c|c|c|c|} \hline 
   		\rule[14pt]{0pt}{0pt}
   	
   		&$|{\rm Re} [\tau]|$
   		&${\rm Im} [\tau]$ 
   		&$|g_u|$
   		&${\rm Arg}[g_u]$ 
   		&$|g_d|$
   		&${\rm Arg} [g_d]$
   		\rule[14pt]{0pt}{0pt}	 \\   		
   		\hline \hline 
   		\rule[14pt]{0pt}{0pt}	
  	range & 0 -- 0.007 &1.013 -- 1.048 & 0 -- 1.396
  		 & $[-\pi,\,\pi]$&0 -- 1.443& $[-\pi,\,\pi]$
  		\rule[14pt]{0pt}{0pt}	\\
  		\hline
   	\end{tabular}
   	\caption{Parameter ranges consistent with the observed CKM mixing angles and   $\delta_{CP}$.}
   	\label{parameters}
   \end{table}
  \begin{figure}[b!]
  	\begin{minipage}[]{0.47\linewidth}
  		\includegraphics[{width=\linewidth}]{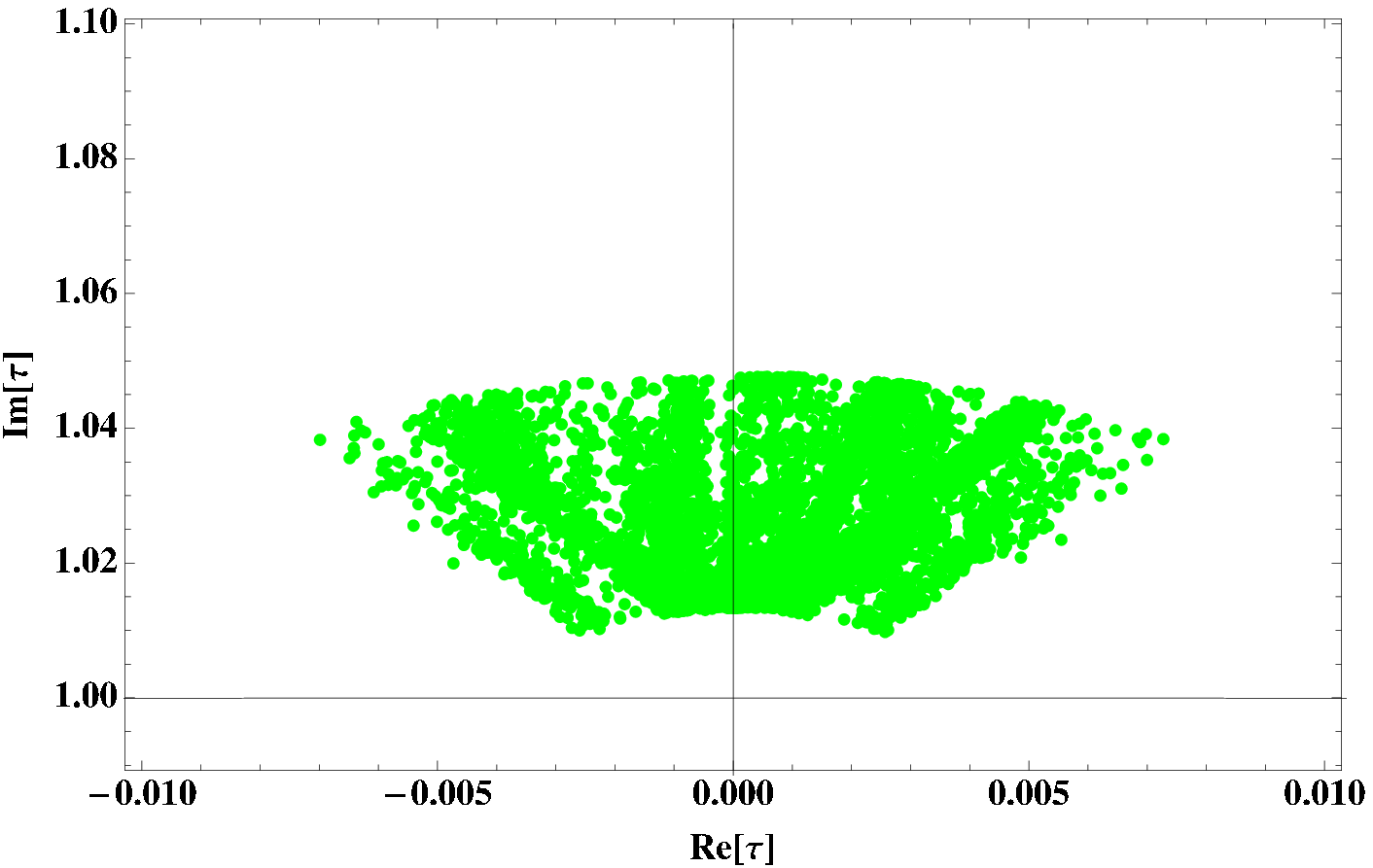}
  		\caption{Allowed region on ${\rm Re} [\tau]$--${\rm Im} [\tau]$
  			plane consistent with three CKM mixing angles and $\delta_{CP}$.
  			The  black curve is the boundary of the fundamental region, $|\tau|=1$.		 
  		}
  	\end{minipage}
  	\hspace{5mm}
  	\begin{minipage}[]{0.47\linewidth}
  		\vspace{-5mm}
  		\includegraphics[{width=\linewidth}]{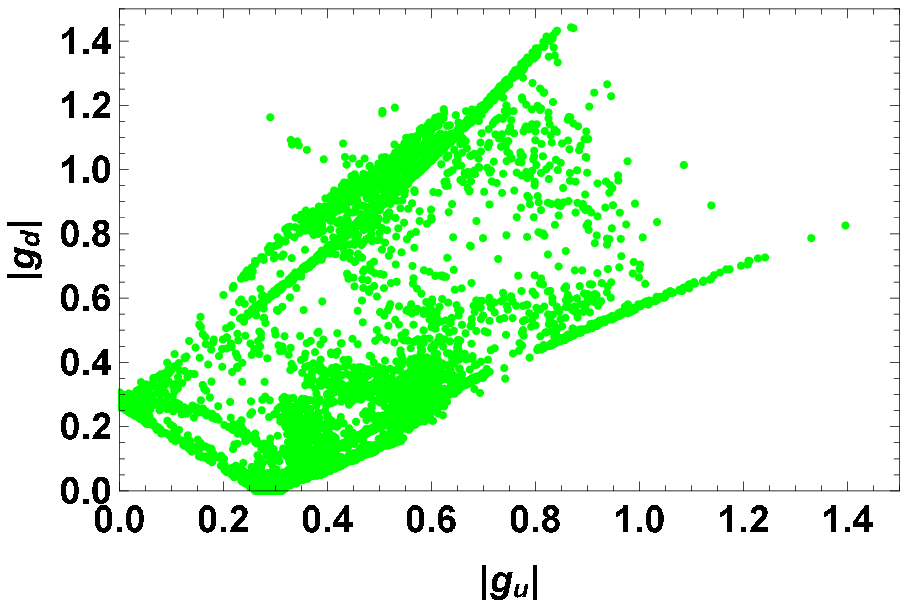}
  		\caption{Allowed region on $|g_u|$--$|g_d|$ plane
  			consistent with three CKM mixing angles and $\delta_{CP}$. 
  		}
  	\end{minipage}
  \end{figure}
   As seen in  Table 2,  the modulus $\tau$ is close to $i$.
     The deviation from $i$ is less than $5\%$.
     The modulus $\tau=i$ is a fixed point in the fundamental region of SL$(2,Z)$ under the transformation $S: \tau\rightarrow -1/\tau$. 
     At  $\tau=i$, CP is not violated as discussed in some works \cite{Baur:2019kwi,Novichkov:2019sqv,Kobayashi:2019uyt}.
     Indeed,  we have succeeded to reproduce  the observed CP violation
     at $\tau$ which is deviated a little bit from  $\tau=i$.
        

   In Fig.\,1,
    we show the plot  of ${\rm Re} [\tau]$ and ${\rm Im} [\tau]$,
    where output points are distributed overall
    in the  $3\,\sigma$ range of the observed CKM elements, $|V_{us}|$, $|V_{cb}|$, $|V_{ub}|$ and $\delta_{CP}$ by choosing relevant $g_u$ and $g_d$.
    It is noticed that  ${\rm Im} [\tau]=1-1.013$ is excluded
     while  $|{\rm Re} [\tau]|$ is very small,  less than $0.007$.

 The fixed point $\tau=i$ is realized 
 if there is a residual symmetry   $\mathbb{Z}_2^{S}=\{ I, S \}$, which is the subgroup of $A_4$.
 Then,  the generator $S$ commutes with $M_{q}^\dagger M_{q}$\,,
 \begin{align}
 \begin{aligned}
 \left [ M_{q}^\dagger M_{q}\, ,\, S \right ]=0 \, .
 \end{aligned}
 \label{commute}
 \end{align}
 Therefore, the mass matrix is expected to be diagonal
 in the diagonal base  $\hat S$.
 However, the eigenvalue $-1$ of $S$ is degenerated, and so
 one pair off diagonal terms appear in  $M_{q}^\dagger M_{q}$.
 We move to the diagonal base  $\hat S$ 
 	by a unitary transformation as:
 	$\hat S= U S U^\dagger$, while the quark mass
 	matrix $M_q$ is transformed as $M_q U^\dagger$.
 	For the diagonal base   $\hat S={\rm diag (-1,-1,1)}$, 
 	the unitary matrix is given as:
 	\begin{align} U=
 	\begin{pmatrix}
 	\frac{2}{\sqrt{6}}&-\frac{1}{\sqrt{6}}&-\frac{1}{\sqrt{6}} \\
 	0&\frac{1}{\sqrt{2}}&\frac{1}{\sqrt{1}}\\
 	\frac{1}{\sqrt{3}}&\frac{1}{\sqrt{3}}&\frac{1}{\sqrt{3}}
 	\end{pmatrix}.
 	\label{unitary}
 	\end{align}
 Indeed, we have  scanned  model parameters
  around $\tau=i$ in the  $\hat S$ base since it is easy to find hierarchical quark mass matrices
  consistent with the observed CKM matrix.
  
    We also show the allowed region of absolute values, $|g_u|$
    and $|g_d|$ in Fig.2.
    It is remarked that  the non-vanishing $|g_u|$ or $|g_d|$ 
    is required to reproduce the CKM elements, however, those are
    at most  of order $1$.
     Indeed, the sum of  $|g_u|$ and $|g_d|$ are larger than $0.25$,
     but smaller than $2.3$.
    
    In Table 3, 
    we  show  typical parameter sets and calculated   CKM parameters.
     Ratios of  $\alpha_q/ \gamma_q$ and $\beta_q/ \gamma_q$ $(q=u,d)$
     correspond to the observed  quark mass hierarchy.
 \begin{table}[h]
 	\centering
 	\begin{tabular}{|c|c|} \hline 
 		&  A  sample set \\
 		\hline\hline
 		\rule[14pt]{0pt}{0pt}
 		$\tau$&   $0.0007 + 1.041\, i$  \\ 
 		\rule[14pt]{0pt}{0pt}
 		$g_u$ & $0.407 - 0.198\, i$ \\
 		\rule[14pt]{0pt}{0pt}
 		$g_d$ & $0.745 + 0.176\, i$ \\
 		\rule[14pt]{0pt}{0pt}
 		$\alpha_u/\gamma_u$ & $ 1.74 \times 10^2$  \\
 		\rule[14pt]{0pt}{0pt}
 		$\beta_u/\gamma_u$ & $1.11\times 10^5$  \\
 		\rule[14pt]{0pt}{0pt}
 		$\alpha_d/\gamma_d$ &  $1.88 \times 10^1$ \\
 		\rule[14pt]{0pt}{0pt}
 		$\beta_d/\gamma_d$ &  $1.70\times 10^{-2}$ \\
 		\rule[14pt]{0pt}{0pt}
 	$|V_{us}|$ &  $0.226 $	\\
 		\rule[14pt]{0pt}{0pt}
 	$|V_{cb}|$ &  $0.0487$	\\
 		\rule[14pt]{0pt}{0pt}
 	$|V_{ub}|$ &  $0.0027$	\\
 		\rule[14pt]{0pt}{0pt}
 	$\delta_{CP}$  & $74.8^\circ$ 	\\
 		\hline
 	\end{tabular}
 	\caption{Numerical values of parameters and output of CKM parameters
 		 at a sample point. }
 	\label{sample}
 \end{table}
We also present the mixing matrices of  up-type quarks and down-type quarks  for  the sample in Table 3
 in order to investigate the flavor structure of each quark mass matrix. 
After moving to the diagonal basis of $ \hat S={\rm diag (-1,-1,1)}$ from the  
original one in Eq.(\ref{STbase}),
  the mixing matrices of up- and down-quarks are given as:
 \begin{align}
\begin{aligned}
V_u&\approx
\begin{pmatrix}
-0.529 & 0.848 &  -0.025 \\
-0.820 - 0.200\, i & -0.514 - 0.125\, i& -0.091  \\ 
-0.089 - 0.017\, i& -0.026 - 0.011\, i& 0.995 - 0.016\, i 
\end{pmatrix} \ , \\
V_d&\approx
\begin{pmatrix}
0.067  & -0.645& -0.761 \\
0.067 & -0.752 + 0.101\, i& 0.643 - 0.086\, i \\
-0.995 - 0.016\, i& -0.094 + 0.005\, i& -0.008 - 0.006\, i
\end{pmatrix}\ .
\end{aligned}
\label{rotation}
\end{align}
The hierarchical flavor structure is partially seen as expected
in the  discussion of Eq.(\ref{unitary}).
After taking account of the degree of freedom of the permutation among
 $A_4$ triplet elements, ($q_1, q_2, q_3$),
we can obtain the observed CKM matrix  $V_{\rm CKM}=V_u^\dagger \, V_d$.
Indeed, the permutation of   $(d_L, s_L, b_L)$ to  $(s_L, b_L, d_L)$,
which corresponds the exchange of columns $(1,2,3)\to (2,3,1)$ in $V_d$,
 gives the observed CKM matrix with keeping  $(u_L, c_L, t_L)$.


 In conclusion, our quark mass matrix
with the $A_4$ modular symmetry 
can successfully  reproduce the CKM mixing matrix completely.
This successful result encourages us to investigate the lepton sector
in the same framework. We discuss the lepton mass matrices
with the $A_4$ modular symmetry in the next section.

\section{Lepton mass matrix in the $A_4$ modular invariance}

  The modular $A_4$ invariance also gives the lepton mass matrix
  in terms of the modulus $\tau$ which is probably common both quarks and leptons if  flavors of quarks and leptons are originated from a same two-dimensional compact space.
  The $A_4$ representations and weights  are assigned for lepton fields relevantly as seen  in Table 4, where
  the left-handed lepton doublets compose a $A_4$ triplet
  and the right-handed charged leptons are $A_4$ singlets.
  Weights of the left-handed leptons and the right-handed charged leptons
    are assigned  like  the quark ones in Table 4 (case I).
   In order to examine the quantitative dependence of our result
   on weights of the right-handed charged leptons, we also consider  other choices of  weights for  the right-handed ones
    as    cases  I\hspace{-.01em}I -- V of Table 4.
    
  
  \begin{table}[h]
  	\centering
  	\begin{tabular}{|c||c|c|c|c|c|c|c|} \hline
  		&$L$&$(e^c,\mu^c,\tau^c)$&$H_u$&$H_d$&$\bf Y_3^{(6)}, 
 Y_{3'}^{(6)}$ & $\bf Y_3^{(4)}, Y_1^{(4)},  Y_{1'}^{(4)}$
   &$\bf Y_3^{(2)}$ \\  \hline\hline 
  		\rule[14pt]{0pt}{0pt}
  		$SU(2)$&$\bf 2$&$\bf 1$&$\bf 2$&$\bf 2$&$\bf 1$& $\bf 1$&$\bf 1$\\
  		$A_4$&$\bf 3$& \bf (1,\ 1$''$,\ 1$'$)&$\bf 1$&$\bf 1$
  		& $\bf 3$& $\bf 3\,, \quad  1\, \quad  1'$ &$\bf 3$\\
  		$-k_I$&$ -2$&I: $(-4, -2, 0)$&0&0& 
  		$k=6$ &$k=4$ &$k=2$ \\
  		& &I\hspace{-.01em}I: $(-4, \ \ 0, \, \ \ 0)$ & & & & &  \\
  & &{\rm I\hspace{-.15em}I\hspace{-.15em}I}: $(0, \ \ 0, \,\ \ 0)$ & & & & &  \\
  		& &I\hspace{-.01em}V: $(-2,\, \ 0, \,\  0)$ & & & & &  \\
  		& &V: $(-2,  \ -2, \,\  -2)$ & & & & &  \\ \hline
  	\end{tabular}	
  \caption{ Assignments of representations and  weights
  	$-k_I$ for MSSM fields and  modular forms.
  	}
  	\label{tb:lepton}
  \end{table}
  Assign the left-handed charged leptons  a $A_4$ triplet
   $L=(e_L, \mu_L,\tau_L)$.
  Let us start with giving the charged lepton mass matrix $M_E$  in terms of
  modular forms of weight $2$, $4$ and $6$ in Eqs.(\ref{Y(2)}),
  	(\ref{weight4}) and (\ref{weight6}) as well as the quark sector.
  For the case I, it is presented as:   
  \begin{align}
  \begin{aligned}
{\rm I :}\quad  M_E=v_d
\begin{pmatrix}
\alpha_e & 0 & 0 \\
0 &\beta_e & 0\\
0 & 0 &\gamma_e
\end{pmatrix} \left [
\begin{pmatrix}
Y_1^{(6)}+g_e Y_1^{'(6)} & Y_3^{(6)} +g_e Y_3^{'(6)} 
& Y_2^{(6)}+g_e Y_2^{'(6)} \\
Y_2^{(4)} & Y_1^{(4)} &  Y_3^{(4)} \\
Y_3^{(2)} &  Y_2^{(2)}&  Y_1^{(2)}
\end{pmatrix}
\right ]_{RL}, 
  \end{aligned}
  \label{ME642}
  \end{align}
where coefficients $\alpha_e$, $\beta_e$ and $\gamma_e$ are real parameters
while $g_e$ is complex one.
For the case I\hspace{-.01em}I, it is given as
\begin{align}
\begin{aligned}
{\rm I\hspace{-.01em}I}: \quad M_E=v_d
\begin{pmatrix}
\alpha_e & 0 & 0 \\
0 &\beta_e & 0\\
0 & 0 &\gamma_e
\end{pmatrix} \left [
\begin{pmatrix}
Y_1^{(6)}+g_e Y_1^{'(6)} & Y_3^{(6)} +g_e Y_3^{'(6)} 
& Y_2^{(6)}+g_e Y_2^{'(6)} \\
Y_2^{(2)} & Y_1^{(2)} &  Y_3^{(2)} \\
Y_3^{(2)} &  Y_2^{(2)}&  Y_1^{(2)}
\end{pmatrix}
\right ]_{RL}.
\end{aligned}
\label{ME622}
\end{align}
On the other hand, a parameter $g_e$ of  the mass matrix disappears
for cases I\hspace{-.15em}I\hspace{-.15em}I, I\hspace{-.01em}V and V.
They are 
\begin{align}
\begin{aligned}
{\rm  I\hspace{-.15em}I\hspace{-.15em}I}: \quad M_E=v_d
\begin{pmatrix}
\alpha_e & 0 & 0 \\
0 &\beta_e & 0\\
0 & 0 &\gamma_e
\end{pmatrix} \left [
\begin{pmatrix}
Y_1^{(2)} & Y_3^{(2)} & Y_2^{(2)} \\
Y_2^{(2)} & Y_1^{(2)} &  Y_3^{(2)} \\
Y_3^{(2)} &  Y_2^{(2)}&  Y_1^{(2)}
\end{pmatrix}
\right ]_{RL}, 
\end{aligned}
\label{ME222}
\end{align}
\begin{align}
\begin{aligned}
{\rm I\hspace{-.01em}V}: \quad M_E=v_d
\begin{pmatrix}
\alpha_e & 0 & 0 \\
0 &\beta_e & 0\\
0 & 0 &\gamma_e
\end{pmatrix} \left [
\begin{pmatrix}
Y_1^{(4)} & Y_3^{(4)} & Y_2^{(4)} \\
Y_2^{(2)} & Y_1^{(2)} &  Y_3^{(2)} \\
Y_3^{(2)} &  Y_2^{(2)}&  Y_1^{(2)}
\end{pmatrix}
\right ]_{RL}, 
\end{aligned}
\label{ME422}
\end{align}
\begin{align}
\begin{aligned}
{\rm V}: \quad M_E=v_d
\begin{pmatrix}
\alpha_e & 0 & 0 \\
0 &\beta_e & 0\\
0 & 0 &\gamma_e
\end{pmatrix} \left [
\begin{pmatrix}
Y_1^{(4)} & Y_3^{(4)} & Y_2^{(4)} \\
Y_2^{(4)} & Y_1^{(4)} &  Y_3^{(4)} \\
Y_3^{(4)} &  Y_2^{(4)}&  Y_1^{(4)}
\end{pmatrix}
\right ]_{RL}, 
\end{aligned}
\label{ME444}
\end{align}
respectively.

Suppose neutrinos to be Majorana particles.
By using the Weinberg operator, the superpotential of the neutrino mass term, $w_\nu$ is  given as:
\begin{align}
w_\nu&=-\frac{1}{\Lambda}(H_u H_u LL{\bf Y_r^{(k)}})_{\bf 1}~,
\label{Weinberg}
\end{align}
where $\Lambda$ is a relevant cut off scale and the $A_4$ singlet component
is extracted.
Since the  left-handed lepton doublet has weight $-2$, the superpotential
is given in terms of  modular forms of weight $4$, ${\bf Y_3^{(4)}}$,
${\bf Y_1^{(4)}}$ and  ${\bf Y_{1'}^{(4)}}$.
By putting the vacuum expectation value of $H_u$ ($v_u$)
and taking $L=(\nu_e, \nu_\mu,\nu_\tau)$ for neutrinos,
 we have
 
\begin{align}
w_\nu &=\frac{v_u^2}{\Lambda}
\left [ 
\begin{pmatrix}
2\nu_e\nu_e-\nu_\mu\nu_\tau-\nu_\tau\nu_\mu\\
2\nu_\tau\nu_\tau-\nu_e\nu_\mu-\nu_\mu\nu_\tau\\
2\nu_\mu\nu_\mu-\nu_\tau\nu_e-\nu_e\nu_\tau
\end{pmatrix} \otimes
{\bf Y_3^{(4)}}  \right . \nonumber \\
& \left .  + \ 
(\nu_e\nu_e+\nu_\mu\nu_\tau+\nu_\tau\nu_\mu)
\otimes g_{\nu 1}{\bf Y_1^{(4)}}
+
(\nu_e\nu_\tau+\nu_\mu\nu_\mu+\nu_\tau\nu_e)
\otimes g_{\nu 2}{\bf Y_{1'}^{(4)}}
\right ]  \nonumber \\
=&\frac{v_u^2}{\Lambda}
\left[(2\nu_e\nu_e-\nu_\mu\nu_\tau-\nu_\tau\nu_\mu)Y_1^{(4)}+
(2\nu_\tau\nu_\tau-\nu_e\nu_\mu-\nu_\mu\nu_e)Y_3^{(4)}
+(2\nu_\mu\nu_\mu-\nu_\tau\nu_e-\nu_e\nu_\tau)Y_2^{(4)}\right .
\nonumber \\
& \left .  + \ 
(\nu_e\nu_e+\nu_\mu\nu_\tau+\nu_\tau\nu_\mu)
g_{\nu 1}{\bf Y_1^{(4)}}
+
(\nu_e\nu_\tau+\nu_\mu\nu_\mu+\nu_\tau\nu_e)
g_{\nu 2}{\bf Y_{1'}^{(4)}}
\right ]   \ , 
\end{align}
where ${\bf Y_3^{(4)}}$, ${\bf Y_1^{(4)}}$ and ${\bf Y_{1'}^{(4)}}$
are given in Eq.\,(\ref{weight4}), and  $g_{\nu 1}$, $g_{\nu 2}$ are complex parameters.
The neutrino mass matrix is written as follows:
\begin{align}
M_\nu=\frac{v_u^2}{\Lambda} \left [
\begin{pmatrix}
2Y_1^{(4)} & -Y_3^{(4)} & -Y_2^{(4)}\\
-Y_3^{(4)} & 2Y_2^{(4)} & -Y_1^{(4)} \\
-Y_2^{(4)} & -Y_1^{(4)} & 2Y_3^{(4)}
\end{pmatrix}
+g_{\nu 1} {\bf Y_{1}^{(4)}  }
\begin{pmatrix}
1 & 0 &0\\ 0 & 0 & 1 \\ 0 & 1 & 0
\end{pmatrix}
+g_{\nu 2} {\bf Y_{1'}^{(4)} }
\begin{pmatrix}
0 & 0 &1\\ 0 & 1 & 0 \\ 1 & 0 & 0
\end{pmatrix}
\right ]_{LL} \ .
\label{neutrinomassmatrix}
\end{align}
 
  Model parameters for cases I and I\hspace{-.01em}I are
    $\alpha_e$, $\beta_e$, $\gamma_e$, $g_e$,  $g_{\nu 1}$ and $g_{\nu 2}$
    apart from the modulus $\tau$
  while those of cases I\hspace{-.15em}I\hspace{-.15em}I,
   I\hspace{-.01em}V and V
  are $\alpha_e$, $\beta_e$, $\gamma_e$,  $g_{\nu 1}$ and $g_{\nu 2}$.
   Parameters $\alpha_e$, $\beta_e$ and  $\gamma_e$ are adjusted 
 by the observed charged lepton masses.
 Therefore, the lepton mixing angles, the Dirac phase  and Majorana phases
 are given by  $g_{\nu 1}$,  $g_{\nu 2}$ $(g_e)$  in addition to the value of $\tau$.  Since  $\tau$ is scanned around $\tau=i$, where  
  the quark CKM matrix is reproduced,
   we expect to get some  predictions in the lepton sector.
  Practically, we scan $\tau$ in the regions of 
    $|{\rm Re} [\tau]|\leq 0.1 $  and ${\rm Im} [\tau]\leq 1.12$,
    which is close to the one of the quark sector
    \footnote{If $\tau$ is scanned as a free parameter in the fundamental region,
     there may be other  regions of $\tau$ which are consistent with  observed lepton mixing angles.}.
    Indeed,  our predictions are almost unchanged even if
     the scanned region of  ${\rm Im} [\tau]$ is enlarged such as 
      ${\rm Im} [\tau]\leq 1.13$.
   

\begin{table}[t!]
	\begin{center}
		\begin{tabular}{|c|c|c|}
			\hline 
			\rule[14pt]{0pt}{0pt}
			\  observable \ &  $3\,\sigma$ range for NH  & $3\,\sigma$ range for IH \\
			\hline 
			\rule[14pt]{0pt}{0pt}
			$\Delta m_{\rm atm}^2$& \ \   \ \ $(2.436$--$ 2.618) \times 10^{-3}\,{\rm eV}^2$ \ \ \ \
			&\ \ $- (2.419$--$2.601) \times 10^{-3}\,{\rm eV}^2$ \ \  \\
			\hline 
			\rule[14pt]{0pt}{0pt}
		$\Delta m_{\rm sol }^2$& $(6.79$--$ 8.01) \times 10^{-5}\,{\rm eV}^2$
			& $(6.79$--$ 8.01)  \times 10^{-5}\,{\rm eV}^2$ \\
			\hline 
			\rule[14pt]{0pt}{0pt}
			$\sin^2\theta_{23}$&  $0.433$--$ 0.609$ & $0.436$--$ 0.610$ \\
			\hline 
			\rule[14pt]{0pt}{0pt}
			$\sin^2\theta_{12}$& $0.275$--$ 0.350$ & $0.275$--$ 0.350$ \\
			\hline 
			\rule[14pt]{0pt}{0pt}
			$\sin^2\theta_{13}$&$0.02044$--$ 0.02435$ & $0.02064$--$0.02457$ \\
			\hline 
		\end{tabular}
		\caption{The $3\,\sigma$ ranges of neutrino  parameters from NuFIT 4.1
			for NH and IH 
			\cite{Esteban:2018azc}. 
		}
		\label{DataNufit}
	\end{center}
\end{table}

\section{Modulus $\tau$ in the lepton mixing}
 We input charged lepton masses in order to constrain the model parameters.
We take Yukawa couplings of charged leptons 
at the GUT scale $2\times 10^{16}$ GeV,  where $\tan\beta=5$ is taken
as well as  quark Yukawa couplings
\cite{Antusch:2013jca, Bjorkeroth:2015ora}:
\begin{eqnarray}
y_e=(1.97\pm 0.024) \times 10^{-6}, \quad 
y_\mu=(4.16\pm 0.050) \times 10^{-4}, \quad 
y_\tau=(7.07\pm 0.073) \times 10^{-3},
\end{eqnarray}
where lepton masses are  given by $m_\ell=y_\ell v_H$ with $v_H=174$ GeV.
We also use 
the  following lepton mixing angles and neutrino mass parameters,
which are given by NuFit 4.1 in Table 5 \cite{Esteban:2018azc}.
Since  there are two possible spectrum of neutrinos masses $m_i$, which are
the normal  hierarchy (NH), $m_3>m_2>m_1$, and the  inverted  hierarchy (IH),
$m_2>m_1>m_3$, we investigate both cases.


Neutrino masses and  
the PMNS matrix $U_{\rm PMNS}$ \cite{Maki:1962mu,Pontecorvo:1967fh} 
 are obtained by diagonalizing 
$M_E^{\dagger} M_E$ and $M_\nu^* M_\nu$.
We also investigate
the sum of three neutrino  masses  $\sum m_i$ in our model since
it is constrained by the recent cosmological data,
\cite{Tanabashi:2018oca,Vagnozzi:2017ovm,Aghanim:2018eyx}.
The effective mass for the $0\nu\beta\beta$ decay is given as follows:
\begin{align}
\langle m_{ee}	\rangle=\left| m_1 c_{12}^2 c_{13}^2+ m_2s_{12}^2 c_{13}^2 e^{i\alpha_{21}}+
 m_3 s_{13}^2 e^{i(\alpha_{31}-2\delta^\ell_{CP})}\right|  \ ,
\end{align}
where $\delta^\ell_{CP}$  is the Dirac phase of leptons,
and $\alpha_{21}$, $\alpha_{31}$ are Majorana phases
 (see Appendix B).

Let us  discuss numerical results for  NH of  neutrino masses in
the case I of Eq.(\ref{ME642}) like the quark sector.
After inputting  charged lepton masses,
parameters $\tau$,  $g_{e}$,
$g_{\nu 1}$ and $g_{\nu 2}$   are constrained by
four observed quantities;
 three mixing angles of leptons
and observed mass ratio $\Delta m_{\rm sol}^2/\Delta m_{\rm atm}^2$.

At first,  we show the allowed region  on the 
${\rm Re} [\tau]$--${\rm Im} [\tau]$ plane in Fig.\,3. Observed three mixing angles of leptons are reproduced at cyan, blue  and red points.
At cyan points, the sum of neutrino masses 
is consistent with  the cosmological upper-bound $120$\,meV.
Red points denote common values of $\tau$ in both quarks  and leptons.
All allowed points are restricted in ${\rm Im} [\tau]\leq 1.08$
and $|{\rm Re} [\tau]|\leq 0.08$, that is around  $\tau=i$.
The red region does not satisfy $\sum m_i\leq 120$meV 
unless it expands to $|{\rm Re} [\tau]|\simeq 0.06$ or 
${\rm Im} [\tau]\simeq 1.07$.
In the  region of $\tau$ in Fig.\,1,  we  discuss
the neutrino masses and the CP violating phase  $\delta_{CP}^\ell$.

\begin{figure}[t!]
	\begin{minipage}[]{0.47\linewidth}
		\includegraphics[{width=\linewidth}]{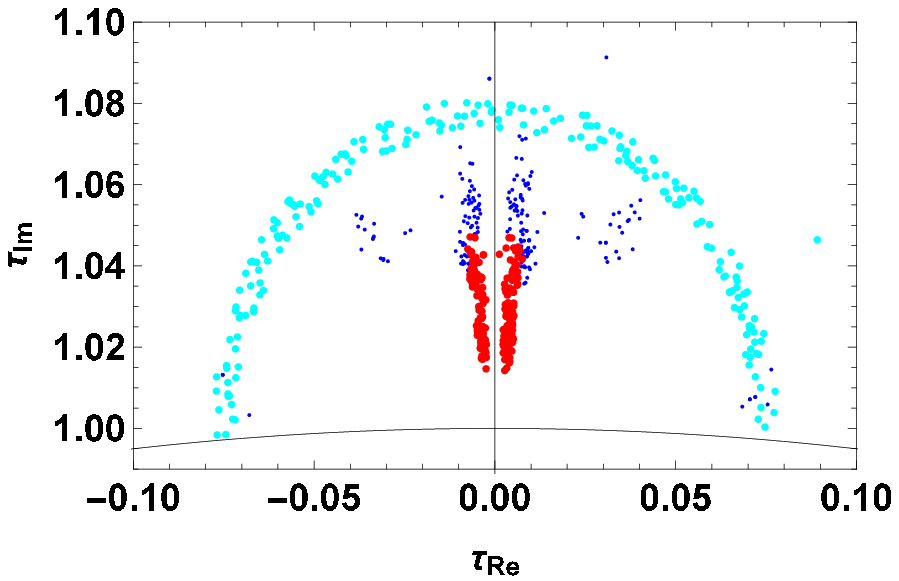}
		\caption{Allowed region on ${\rm Re} [\tau]$--${\rm Im} [\tau]$
			plane  for NH  in the case I of $M_E$.
			Observed  mixing angles are reproduced
			at cyan, blue and red  points. 
			At cyan points, the sum of neutrino masses 
			is below $120$\,meV. 
			The  solid curve is the boundary of the fundamental region, $|\tau|=1$.  }	
	\end{minipage}
	\hspace{5mm}
	\begin{minipage}[]{0.47\linewidth}
		\vspace{-5mm}
		\includegraphics[{width=\linewidth}]{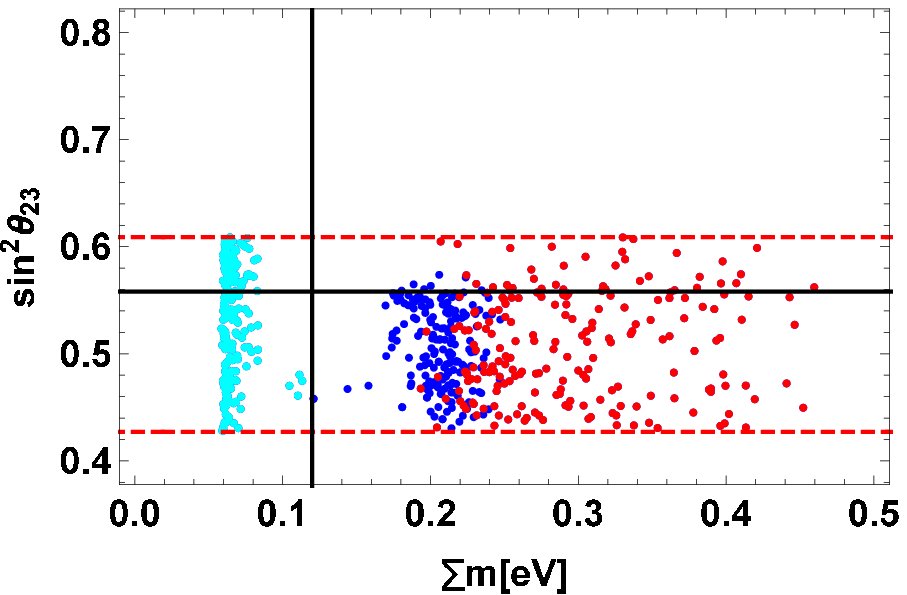}
		\caption{Allowed region on $\sum m_i$--$\sin^2\theta_{23}$ plane,
			where horizontal solid line denotes observed best-fit value,
			red dashed-lines denote
			the bound  of  $3\sigma$ interval,
			and  vertical  line is the cosmological bound,
			for NH  in the case I of $M_E$.
			Color of  points correspond to   $\tau$ in Fig.3.}
	\end{minipage}
\end{figure}

  \begin{figure}[b!]
  	\begin{minipage}[]{0.47\linewidth}
  			\vspace{-5mm}
  	\includegraphics[{width=\linewidth}]{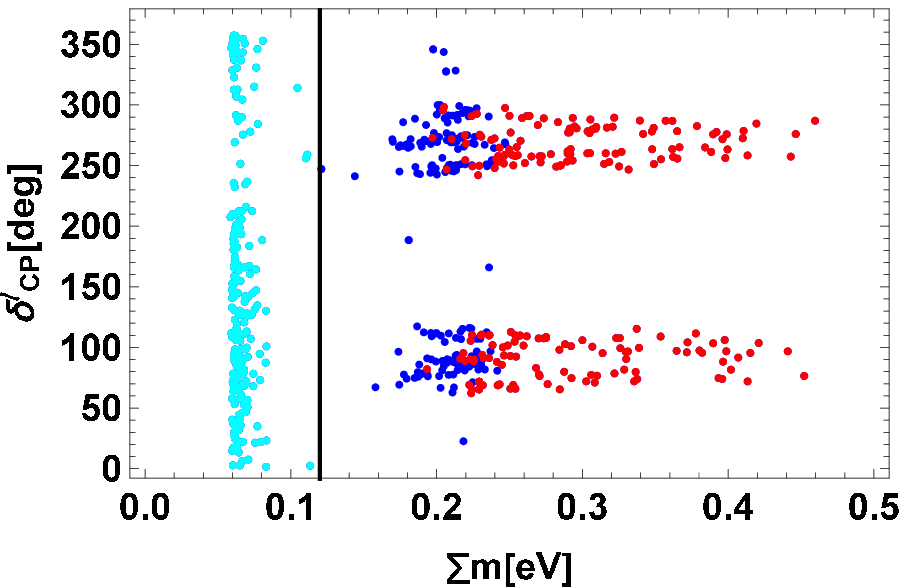}
  	\caption{Allowed region on the  $\sum m_i$--$\delta_{CP}^\ell$ plane,
  		where the vertical  line is the cosmological bound,
  		for NH  in the case I of $M_E$.
  		Colors of points correspond to   $\tau$ in Fig.3.}
  	\end{minipage}
  	\hspace{5mm}
  	\begin{minipage}[]{0.47\linewidth}
  		\vspace{-5mm}
  \includegraphics[{width=\linewidth}]{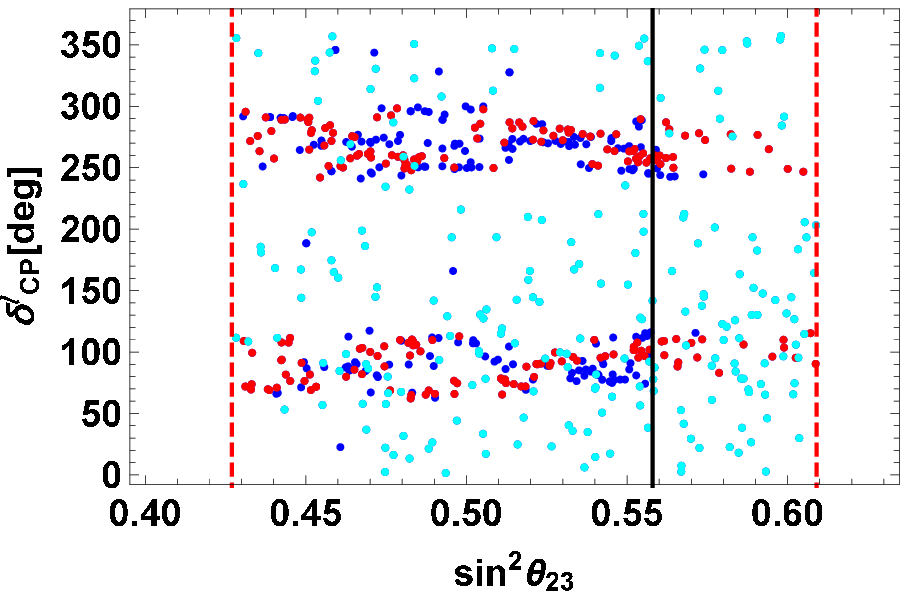}
  \caption{The correlation of $\sin^2\theta_{23}$ and $\delta_{CP}^\ell$,
  	where the  solid  line denotes observed best-fit value and red dashed-lines denote
  	the bound  of  $3\sigma$ interval,
  	for NH  in the case I of $M_E$.}
  	\end{minipage}
  \end{figure}
 

We show the allowed region on  the $\sum m_i$--$\sin^2\theta_{23}$ plane
 in Fig.\,4,
where colors (cyan, blue and red) of points correspond to points of  $\tau$ in Fig.\,3.
The sum of neutrino masses is  constrained by
 the cosmological upper-bound as seen in  Fig.\,4.
The minimal cosmological model, ${\rm \Lambda CDM}+\sum m_i$,
provides a tight bound for the sum of neutrino masses,  $\sum m_i<120$\,meV
\cite{Vagnozzi:2017ovm,Aghanim:2018eyx} although it becomes  weaker when the data are analysed in the context of extended cosmological models \cite{Tanabashi:2018oca}.
It is noticed that red points are in $\sum m_i\geq 190$meV in Fig.\,4.
If these points will be completely exculded by the robust cosmological upper-bound of  the sum of neutrino masses in the near future, the common region of $\tau$ between quarks and leptons vanishes.
The calculated  $\sin^2\theta_{23}$ is distributed overall
in the  $3\,\sigma$ range of NuFIT 4.1 \cite{Esteban:2018azc}
below  $\sum m_i=120$\,meV.

We show the allowed region 
on the  $\sum m_i$--$\delta_{CP}^\ell$ plane in Fig.\,5.
In the region of red points, $\delta_{CP}^\ell$ is predicted to be in 
the restricted ranges,  
$60^\circ$--$120^\circ$ and  $240^\circ$--$300^\circ$.
 Below   $\sum m_i=120$\,meV (cyan points),
  $\delta_{CP}^\ell$
  is allowed in $[0,\, 2\pi]$.
 In Fig.\,6,  we  plot $\delta_{CP}^\ell$ versus $\sin^2\theta_{23}$  
  in order to see their correlation. 
    It is found that there is no distinct correlation 
    between   them.
    Around  the best fit point of $\sin^2\theta_{23}$, 
    the predicted $\delta_{CP}^\ell$ is also in $[0,\,2\pi]$
    below   $\sum m_i=120$\,meV (cyan points).

   It is also noted that 
   the effective mass of the $0\nu\beta\beta$ decay
  $\langle m_{ee}\rangle$
  is predicted in  $1.5$--$28$\,meV below   $\sum m_i=120$\,meV.

%


 \begin{table}[h!]
 	\centering
 	\begin{tabular}{|c|c|} \hline 
 		\rule[14pt]{0pt}{0pt}	  & A sample set \\ \hline \hline 
 		\rule[14pt]{0pt}{0pt}	
 		$\tau$&   $0.028 + 1.075 \, i$  \\ 
 		\rule[14pt]{0pt}{0pt}
 		$g_{\nu 1}$ &$-1.156 - 0.373\, i$ \\
 		\rule[14pt]{0pt}{0pt}
 		$g_{\nu 2}$  &  $-0.617 - 0.441\, i$ \\
 		\rule[14pt]{0pt}{0pt}
 		$g_{e}$ & $-0.736 - 0.430\, i$ \\
 		\rule[14pt]{0pt}{0pt}
 		$\alpha_e/\gamma_e$ & $11.6$  \\
 		\rule[14pt]{0pt}{0pt} 
 		$\beta_e/\gamma_e$ &  $3.92\times 10^{-3}$  \\
 		\rule[14pt]{0pt}{0pt}
 		$\sin^2\theta_{12}$ & $0.314 $	\\
 		\rule[14pt]{0pt}{0pt}
 		$\sin^2\theta_{23}$ &  $0.563$	\\
 		\rule[14pt]{0pt}{0pt}
 		$\sin^2\theta_{13}$ &  $0.0233$	\\
 		\rule[14pt]{0pt}{0pt}
 		$\delta_{CP}^\ell$ &  $278^\circ$ 	\\
 		\rule[14pt]{0pt}{0pt}
 		$\sum m_i$ &  $72.2$\,meV 	\\
 		\rule[14pt]{0pt}{0pt}
 		$\langle m_{ee} \rangle$ &  $8.0$\,meV 	\\
 		\hline
 	\end{tabular}
 	\caption{Numerical values of parameters and output of PMNS parameters
 		at a sample point.}
 	\label{samplelepton}
 \end{table}
In Table 6, 
we  show  a typical parameter set and the output of  observables, 
which is chosen among cyan points (below   $\sum m_i=120$\,meV). 
Ratios of  $\alpha_e/ \gamma_e$ and $\beta_e/ \gamma_e$ 
correspond to  the observed  charged lepton mass hierarchy.

As discussed in the quark sector,
Eqs.\,(\ref{commute}) and (\ref{unitary}), we have  scanned  model parameters
around $\tau=i$ in the diagonal base of the generator $S$,
where eigenvalues are ${\rm diag (1,-1,-1)}$ in this case.
The unitary matrix corresponding to Eq.(\ref{unitary}) 
is obtained by its permutation of rows $1 \to  3 \to  2 \to  1$.
Thanks to this base, it is easy to find hierarchical mixing matrices of the charged lepton.
In this base, we present the mixing matrices of   charged leptons and
neutrinos for a sample of Table 6 as:
\begin{align}
\begin{aligned}
U_\ell&\approx
\begin{pmatrix}
0.627 + 0.759 \,i & -0.017 - 0.077\, i& -0.159 - 0.005\, i\\
	 -0.107- 0.064\, i& -0.560 + 	0.237\, i& -0.782 - 0.063\, i\\
	 -0.108- 0.064\, i& 0.221 - 0.759 \,i& -0.404 + 0.444\, i\\
\end{pmatrix} \ , \\
U_\nu&\approx
\begin{pmatrix}
-0.101 - 0.856\, i & -0.190 + 0.470 \,i& 0.008 - 0.001\, i\\
-0.488 - 0.107\, i & -0.640 - 0.557 \, i& -0.162 + 0.068 \,i\\
 0.078 + 0.035\, i&	0.081 + 0.131 \,i& -0.978 + 0.116 \, i
\end{pmatrix} \ .
\end{aligned}
\label{rotationL}
\end{align}
The PMNS matrix is given by $U_{\rm PMNS}=U_\ell^\dagger \, U_\nu$.
It is noticed that the  large $\theta_{23}$ comes from  the charged lepton mass matrix while the large $\theta_{12}$ comes from  the neutrino one.

 In order to study the quantitative dependence of our result
on weights of the right-handed charged leptons, we discuss other cases of
 those weights in Table 4, in which
  the mass matrix of the charged lepton is different from the quark one.
 Let us begin to examine the  case  I\hspace{-.01em}I of Table 4,
  which presents the charged lepton mass matrix in Eq.(\ref{ME622}).
  Indeed, we obtain the common region of $\tau$ for quarks and leptons in this case.
  Let us  show the allowed region  
  ${\rm Re} [\tau]$--${\rm Im} [\tau]$ plane in Fig.\,7. Observed three mixing angles of leptons are reproduced at cyan, blue and red points.
  At cyan points, the sum of neutrino masses 
  is below the cosmological upper-bound $120$\,meV. 
  Red points denote common values of $\tau$ in both quarks  and leptons.
  The red region does not satisfy $\sum m_i\leq 120$meV 
   unless it expands to $|{\rm Re} [\tau]|\simeq 0.03$ and ${\rm Im} [\tau]\simeq 1.06$.

 \begin{figure}[t!]
 	\begin{minipage}[]{0.47\linewidth}
 		\includegraphics[{width=\linewidth}]{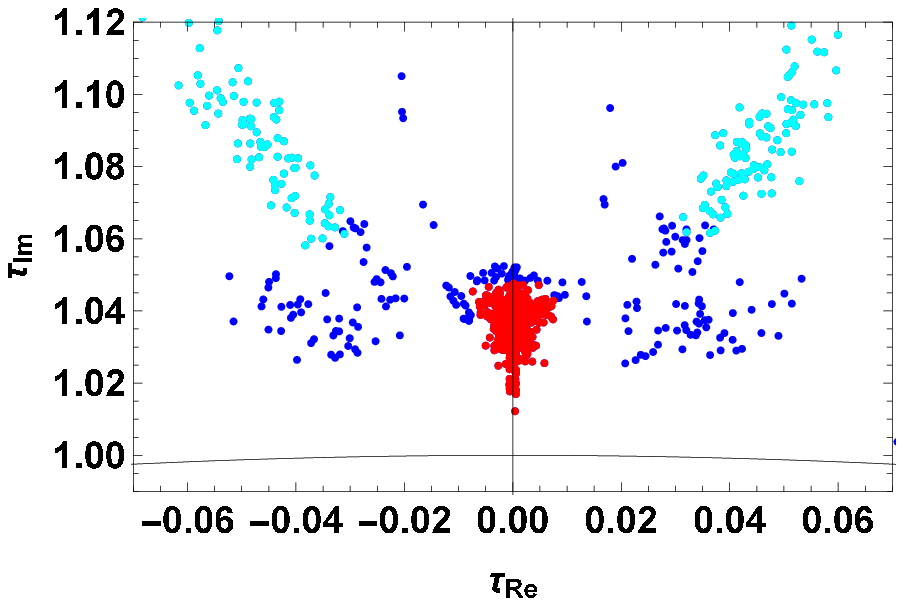}
 		\caption{Allowed region on ${\rm Re} [\tau]$--${\rm Im} [\tau]$
 			plane  for NH  in the case   I\hspace{-.01em}I of $M_E$.
 			Observed  mixing angles are reproduced
 			at cyan, blue and red  points. 
 			At cyan points, the sum of neutrino masses 
 			is below $120$\,meV.
 			The  solid curve is the boundary of the fundamental region, $|\tau|=1$.  }	
 	\end{minipage}
 	\hspace{5mm}
 	\begin{minipage}[]{0.47\linewidth}
 		\vspace{-5mm}
 		\includegraphics[{width=\linewidth}]{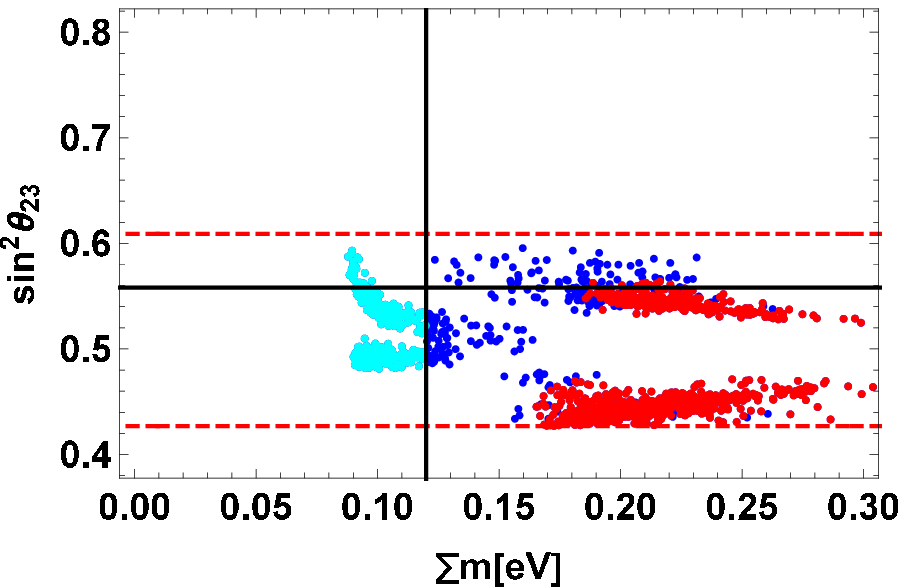}
 	\caption{Allowed region on  $\sum m_i$--$\sin^2\theta_{23}$ plane,
 			where  horizontal solid line denotes observed best-fit one, 
 			    red dashed-lines denote
 			the bound  of  $3\sigma$ interval,
 			and vertical  line is the cosmological  bound,
 			for NH  in the case I\hspace{-.01em}I of $M_E$.
 			Colors of points correspond to $\tau$ in Fig.7.}
 	\end{minipage}
 \end{figure}
We show the allowed region on  the $\sum m_i$--$\sin^2\theta_{23}$ plane
in Fig.\,8,
where colors (cyan, blue and red) of points correspond to points of  $\tau$ in Fig.8.
The sum of neutrino masses is  constrained by
the cosmological upper-bound as seen in   Fig.\,7.
It is noticed that red points are in $\sum m_i\geq 165$meV.
If these points will be completely excluded by the robust cosmological upper-bound of  the sum of neutrino masses in the near future, the common region of $\tau$ between quarks and leptons vanishes.
Below $\sum m_i= 120$meV, $\sin^2\theta_{23}$ is predicted in the  range of $0.48$--$0.60$.
\vskip 0.5 cm
\begin{figure}[h!]
	\begin{minipage}[]{0.47\linewidth}
		\vspace{-5mm}
		\includegraphics[{width=\linewidth}]{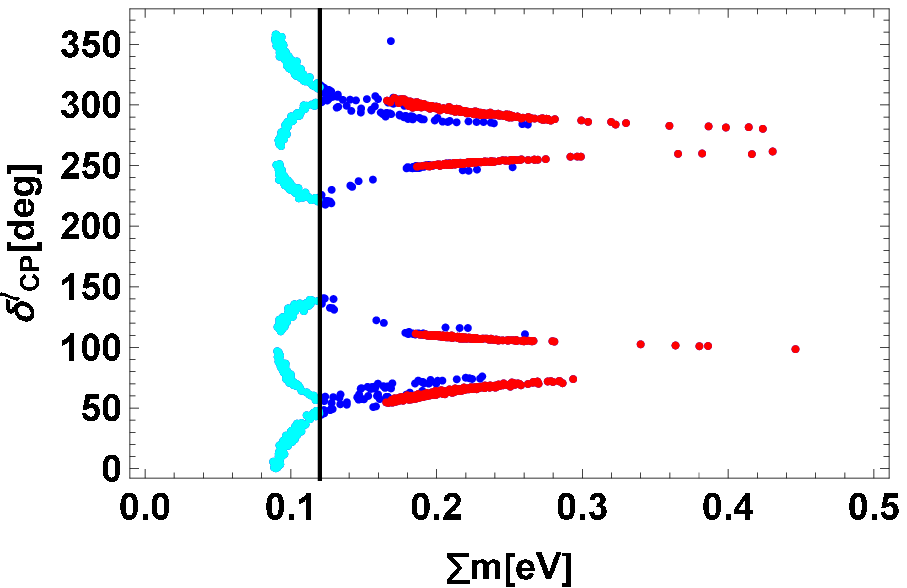}
		\caption{Allowed region on   $\sum m_i$--$\delta_{CP}^\ell$ plane,
			where the vertical  line is the cosmological  bound,
			for NH  in the case I\hspace{-.01em}I of $M_E$.
			Colors of points correspond to  $\tau$ in Fig.7.}
	\end{minipage}
	\hspace{5mm}
	\begin{minipage}[]{0.47\linewidth}
		\vspace{-5mm}
		\includegraphics[{width=\linewidth}]{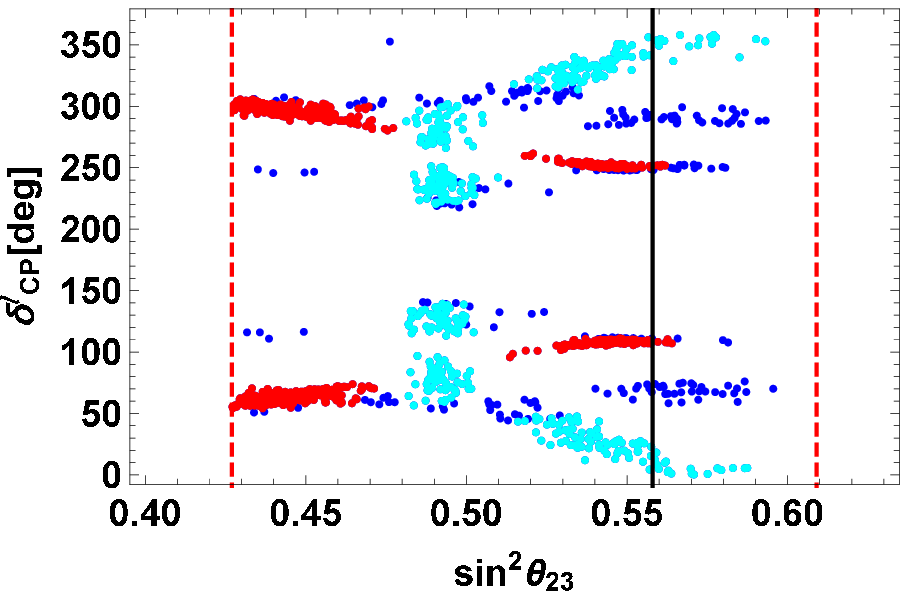}
		\caption{The correlation of $\sin^2\theta_{23}$ and $\delta_{CP}^\ell$,
			where the  solid line denotes observed best-fit value and     red dashed-lines denote
			the bound  of  $3\sigma$ interval,
			for NH  in the case I\hspace{-.01em}I of $M_E$.}
	\end{minipage}
\end{figure}

 We show the allowed region 
 on the  $\sum m_i$--$\delta_{CP}^\ell$ plane in Fig.\,9.
 In the region of red points, $\delta_{CP}^\ell$ is predicted to be in 
 the restricted ranges, 
 $55^\circ$--$75^\circ$, $100^\circ$--$110^\circ$,
 $250^\circ$--$260^\circ$ and $285^\circ$--$305^\circ$.
 Below   $\sum m_i=120$\,meV,
 $\delta_{CP}^\ell$ is rather wide ranges in 
 $0^\circ$--$100^\circ$,  $110^\circ$--$135^\circ$,
  $225^\circ$--$250^\circ$ and $260^\circ$--$360^\circ$.
 We plot  $\delta_{CP}^\ell$ versus $\sin^2\theta_{23}$
  in  Fig.\,10.
  It is found a clear correlation between  them in contrast with the case I.
  At the best fit point of $\sin^2\theta_{23}$, 
  $\delta_{CP}^\ell$ is predicted to be $0^\circ$--$20^\circ$ and $340^\circ$--$360^\circ$ below   $\sum m_i=120$\,meV (cyan points).

   We can also predict
   the effective mass of the $0\nu\beta\beta$ decay
   $\langle m_{ee}\rangle$.
   It is  in $16$--$31$\,meV
    below $\sum m_i=120$\,meV. 
 
\begin{figure}[h!]
	\begin{minipage}[]{0.47\linewidth}
		\includegraphics[{width=\linewidth}]{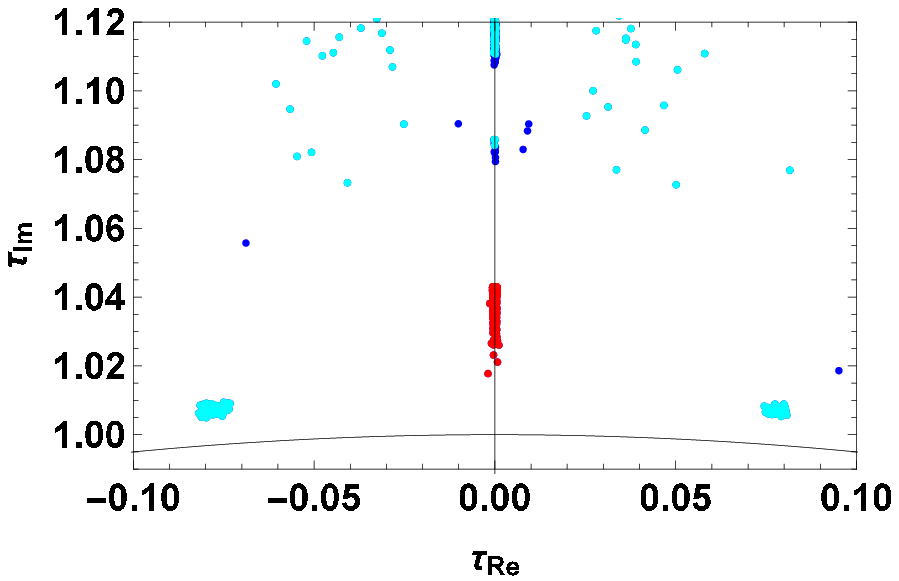}
		\caption{Allowed region on ${\rm Re} [\tau]$--${\rm Im} [\tau]$
			plane for NH  in the case  I\hspace{-.15em}I\hspace{-.15em}I of $M_E$.
			Observed  mixing angles  are reproduced
			at cyan, blue and red  points. 
			At cyan points, the sum of neutrino masses 
			is below $120$\,meV.
			The  solid curve is the boundary of the fundamental region, $|\tau|=1$.  }	
	\end{minipage}
	\hspace{5mm}
	\begin{minipage}[]{0.47\linewidth}
		\vspace{-5mm}
		\includegraphics[{width=\linewidth}]{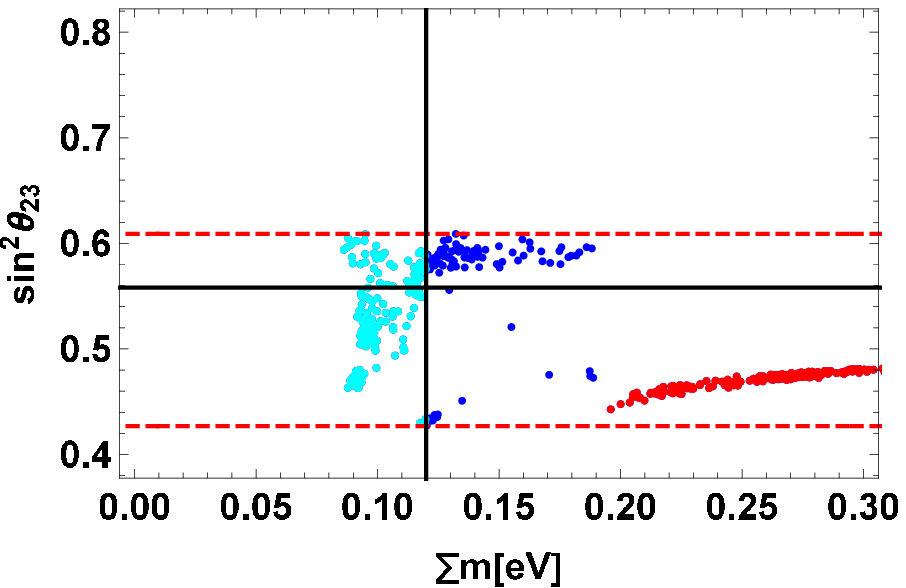}
		\caption{Allowed region on   $\sum m_i$--$\sin^2\theta_{23}$ plane,
			where horizontal solid line denotes observed best-fit one,  red dashed-lines denote
			the bound  of  $3\sigma$ interval,
			and vertical  line is the cosmological bound,
			for NH  in the case  I\hspace{-.15em}I\hspace{-.15em}I of $M_E$.
			Colors of points correspond to   $\tau$ in Fig.11.}
	\end{minipage}
\end{figure}

As seen in the charged lepton mass matrix of Eqs.\,(\ref{ME642}) and (\ref{ME622}), there is a complex parameter $g_e$ in addition to $g_{\nu 1}$ and $g_{\nu 2}$ of the neutrino sector
for cases I and  I\hspace{-.01em}I.
This additional parameter $g_e$ gives the  disadvantage to 
predict the CP violating phase.
Indeed,   $\delta_{CP}^\ell$ 
is predicted in the wide range for  cases I and I\hspace{-.01em}I 
as seen in Figs.\,5 and 9.
In order to improve the  predictability of the model,
cases I\hspace{-.15em}I\hspace{-.15em}I, I\hspace{-.01em}V and V 
provide  attractive mass matrices 
\footnote{These textures are not viable  in the quark sector
	because  only one complex parameter $\tau$ cannot reproduce observed four CKM elements.} .

The mass matrix of Eq.\,(\ref{ME222}) (case I\hspace{-.15em}I\hspace{-.15em}I)
is the simplest one although  this texture is excluded in the quark sector as discussed in  Eq.\,(\ref{matrixSM}). 
Remarkably, there exists a common region of $\tau$ for quarks and leptons in this case.
Let us  show the allowed region  on the
${\rm Re} [\tau]$--${\rm Im} [\tau]$ plane in Fig.\,11. Observed three mixing angles of leptons are reproduced at cyan, blue and red points.
The common region  of $\tau$ in both quarks  and leptons
are presented by red points.  At cyan points, the sum of neutrino masses 
is below the cosmological upper-bound $120$\,meV.


We show the allowed region on  the $\sum m_i$--$\sin^2\theta_{23}$ plane
in Fig.\,12, where colors (cyan, blue and red) of points correspond to points of  $\tau$ in Fig.\,11. 
It is found  red points to be  larger than  $\sum m_i=195$meV.
The common region of $\tau$ in quarks and leptons vanishes 
if the cosmological upper-bound of  the sum of neutrino masses is the robust one.
Below $\sum m_i= 120$meV (cyan points), $\sin^2\theta_{23}$ is predicted in the range of $0.46$--$0.60$.
\begin{figure}[h!]
	\vskip 0.5 cm
	\begin{minipage}[]{0.47\linewidth}
		\vspace{-5mm}
		\includegraphics[{width=\linewidth}]{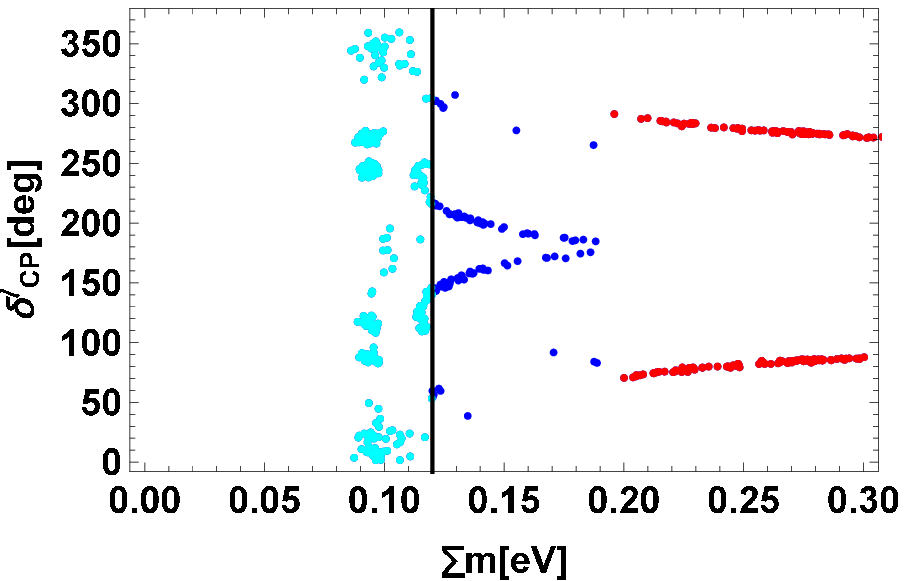}
		\caption{Allowed region on the  $\sum m_i$--$\delta_{CP}^\ell$ plane,
			where the vertical  line is the cosmological bound,
			for NH  in the case  I\hspace{-.15em}I\hspace{-.15em}I of $M_E$.
			Colors of points correspond to points of  $\tau$ in Fig.11.}
	\end{minipage}
	\hspace{5mm}
	\begin{minipage}[]{0.47\linewidth}
		\vspace{-5mm}
		\includegraphics[{width=\linewidth}]{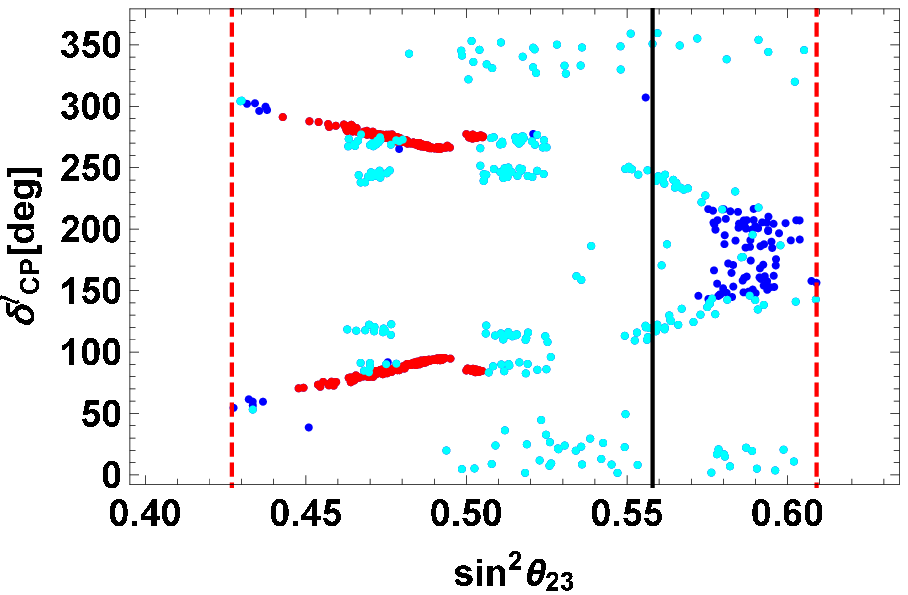}
		\caption{The correlation of $\sin^2\theta_{23}$ and $\delta_{CP}^\ell$,
			where the  black line denotes observed best-fit value and  red dashed-lines denote
			the bound  of  $3\sigma$ interval,
			for NH  in the case  I\hspace{-.15em}I\hspace{-.15em}I of $M_E$.}
	\end{minipage}
\end{figure}

 We show the allowed region 
on the  $\sum m_i$--$\delta_{CP}^\ell$ plane in Fig.\,13.
In the region of red points, $\delta_{CP}^\ell$ is predicted to be in 
the restricted ranges, 
$70^\circ$--$90^\circ$ and  $270^\circ$--$290^\circ$.
However, below   $\sum m_i=120$\,meV,
the predicted region of  $\delta_{CP}^\ell$ is expanded.

We plot  $\delta_{CP}^\ell$ versus $\sin^2\theta_{23}$ in  Fig.\,14.
It is found a clear correlation between  them as well as the case I\hspace{-.01em}I.
At the best fit point of $\sin^2\theta_{23}$, 
$\delta_{CP}^\ell$ is predicted to be $110^\circ$--$125^\circ$ and $235^\circ$--$250^\circ$ below   $\sum m_i=120$\,meV (cyan points),
 which is remarkably different from the ones in cases I and I\hspace{-.01em}I.
We can also predict
the effective mass of the $0\nu\beta\beta$ decay
$\langle m_{ee}\rangle$.
It is  in $15$--$30$\,meV below $\sum m_i=120$\,meV. 

\begin{figure}[h!]
	\begin{minipage}[]{0.47\linewidth}
		\includegraphics[{width=\linewidth}]{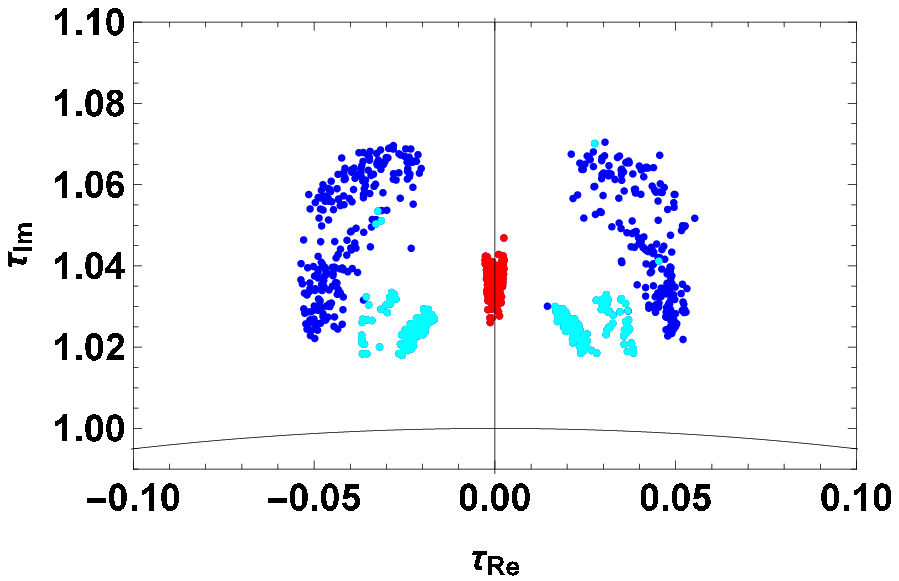}
		\caption{Allowed region on ${\rm Re} [\tau]$--${\rm Im} [\tau]$
			plane for NH  in the case  I\hspace{-.01em}V of $M_E$.
			Observed  mixing angles  are reproduced
			at cyan, blue and red  points. 
			At cyan points, the sum of neutrino masses 
			is below $120$\,meV.
			The  solid curve is the boundary of the fundamental region, $|\tau|=1$.  }	
	\end{minipage}
	\hspace{5mm}
	\begin{minipage}[]{0.47\linewidth}
		\vspace{-5mm}
		\includegraphics[{width=\linewidth}]{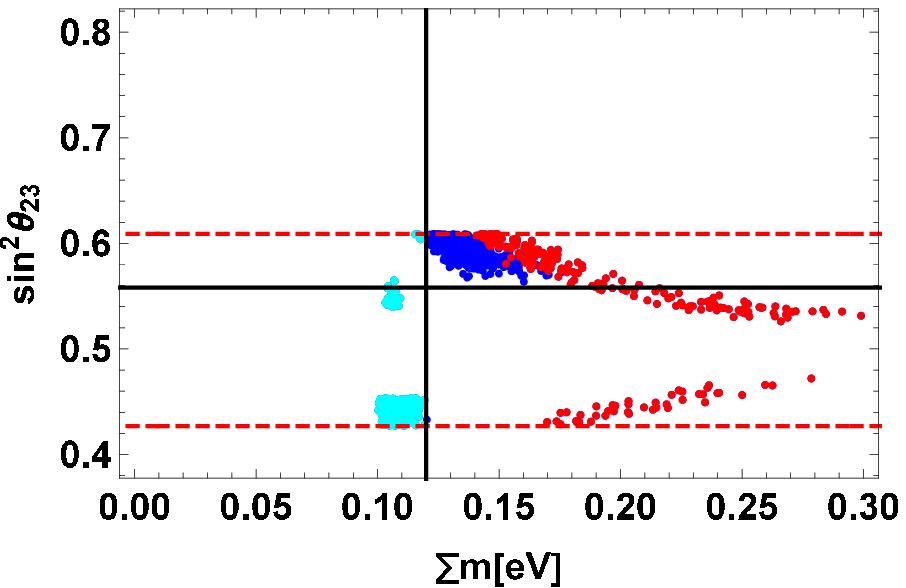}
		\caption{Allowed region on $\sum m_i$--$\sin^2\theta_{23}$ plane,
			where  horizontal solid line denotes observed best-fit value,      dashed-lines denote
			the bound  of  $3\sigma$ interval,
			and  vertical  line is the cosmological  bound,
			for NH  in the case  I\hspace{-.01em}V of $M_E$.
			Colors of points correspond to   $\tau$ in Fig.15.}
	\end{minipage}
\end{figure}

For the case  I\hspace{-.01em}V  in  Eq.\,(\ref{ME422}),
where the charged lepton mass matrix is composed of both weight 2 and 4 modular forms contrast to the case I\hspace{-.15em}I\hspace{-.15em}I, we  obtain the common region of $\tau$ in  quarks and leptons.
Let us  show the allowed region  
${\rm Re} [\tau]$--${\rm Im} [\tau]$ plane in Fig.\,15. Observed three mixing angles of leptons are reproduced at cyan, blue and red points.
At cyan points, the sum of neutrino masses 
is below the cosmological upper-bound $120$\,meV. 
Red points denote common values of $\tau$ in both quarks  and leptons.
As seen in Fig.\,15,  cyan points are not so far away from the red region.  

\begin{figure}[t!]
	\begin{minipage}[]{0.47\linewidth}
		\includegraphics[{width=\linewidth}]{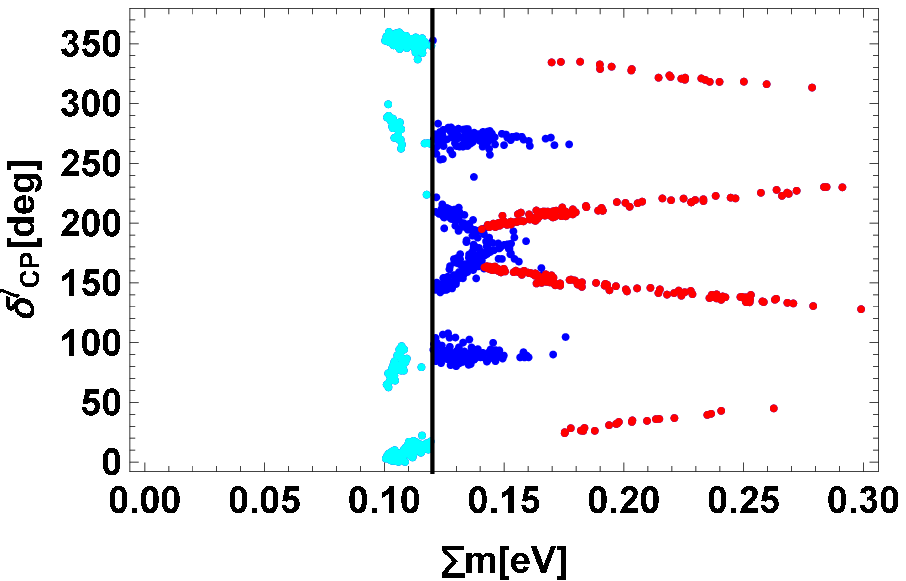}
		\caption{Allowed region on the  $\sum m_i$--$\delta_{CP}^\ell$ plane,
			where the vertical  line is the cosmological bound,
			for NH  in the case  I\hspace{-.01em}V of $M_E$.
			Colors of points correspond to  $\tau$ in Fig.15.}
	\end{minipage}
	\hspace{5mm}
	\begin{minipage}[]{0.47\linewidth}
		\vspace{-5mm}
		\includegraphics[{width=\linewidth}]{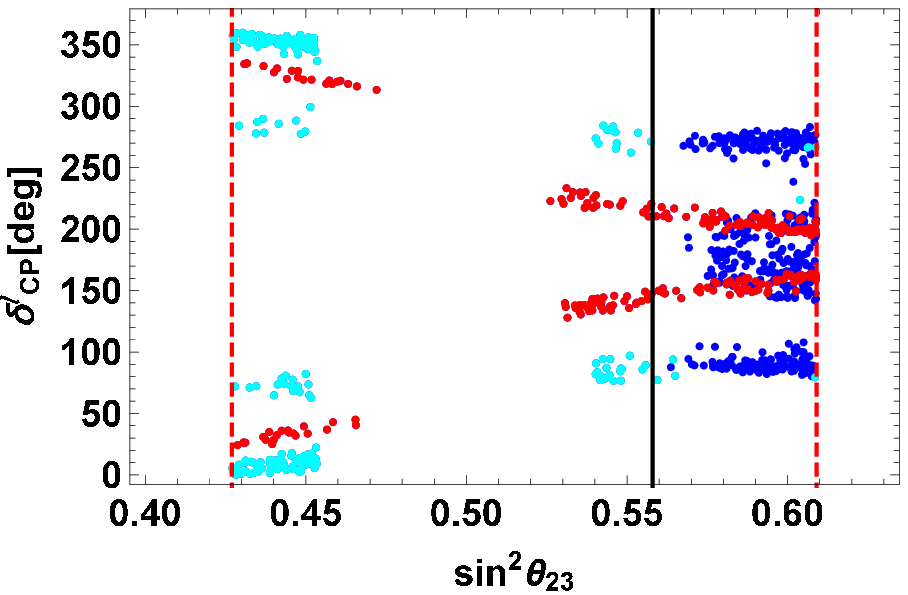}
		\caption{The correlation of $\sin^2\theta_{23}$ and $\delta_{CP}^\ell$,
			where the  solid line denotes observed best-fit value and     red dashed-lines denote
		the bound  of  $3\sigma$ interval,
			for NH  in the case  I\hspace{-.01em}V of $M_E$.}
	\end{minipage}
\end{figure}

We show the allowed region on  the $\sum m_i$--$\sin^2\theta_{23}$ plane
in Fig.\,16, where colors (cyan, blue and red) of points correspond to points of  $\tau$ in Fig.15. 
It is found  red points to be  larger than  $\sum m_i=140$meV.
The common region of $\tau$ in quarks and leptons vanishes 
under $\sum m_i=120$meV.
If the cosmological upper-bound of  the sum of neutrino masses is allowed
up to   $\sum m_i= 160$meV \cite{Tanabashi:2018oca}, 
$\sin^2\theta_{23}$ is predicted  
to be $0.58$--$0.61$ at the common $\tau$ of quarks and leptons.

We show the allowed region 
on the  $\sum m_i$--$\delta_{CP}^\ell$ plane in Fig.\,17.
 The predicted $\delta_{CP}^\ell$ is  in 
 $150^\circ$--$160^\circ$ and  $200^\circ$--$210^\circ$
   at the common $\tau$ of quarks and leptons.
However, below   $\sum m_i=120$\,meV,
the predicted $\delta_{CP}^\ell$ (cyan points) is 
$0^\circ$--$20^\circ$, $60^\circ$--$100^\circ$,
$260^\circ$--$300^\circ$ and  $340^\circ$--$360^\circ$.

We plot $\delta_{CP}^\ell$ versus $\sin^2\theta_{23}$
in  Fig.\,18.
It is found a clear correlation between  them as well as  cases I\hspace{-.01em}I and  I\hspace{-.15em}I\hspace{-.15em}I.
At the best fit point of $\sin^2\theta_{23}$, 
$\delta_{CP}^\ell$ is predicted to be $90^\circ$ and $270^\circ$
below   $\sum m_i=120$\,meV (cyan points).

Finally, the effective mass of the $0\nu\beta\beta$ decay
$\langle m_{ee}\rangle$
 is  in $22$--$30$\,meV
below $\sum m_i=120$\,meV. 


For the case V  in Eq.(\ref{ME444}),
where the charged lepton mass matrix is composed of only weight 4 modular forms, there is no common region of $\tau$ for quarks and leptons.
 Observed three mixing angles of leptons are not reproduced
 unless the sum of neutrino masses  is larger than $450$\,meV,
which is far away from   the cosmological upper-bound $120$\,meV. 
Therefore, we omit to show numerical results for this case.

In Table 7, we summarize characteristic  results of cases I--V 
including  ones of  IH.
In our numerical calculations, we have not included  the RGE effects
in the lepton mixing angles and neutrino mass ratio
$\Delta m_{\rm sol}^2/\Delta m_{\rm atm}^2$.
We suppose that those corrections  are very small between 
the electroweak  and GUT scales
for NH of neutrino masses.
This assumption is  justified very well in the case of $\tan\beta\leq 5$
as far as the  sum of neutrino masses is less than a few hundred meV
 \cite{Criado:2018thu,Haba:1999fk}.
\begin{table}[h]
	\centering
	\begin{tabular}{|c|c|c|c|c|c|c|} \hline
		\rule[14pt]{0pt}{0pt}
	Cases& & I & I\hspace{-.01em}I &I\hspace{-.15em}I\hspace{-.15em}I&I \hspace{-.01em}V&V \\ \hline\hline
		\rule[16pt]{0pt}{0pt}
		common $\tau$	&NH &   $\bigcirc$ & $\bigcirc$&$\bigcirc$
		& $\bigcirc$& $\times$\\  
		\rule[14pt]{0pt}{0pt}
		of quarks/leptons & IH &$\bigcirc$&$\times$&$\bigcirc$&$\times$
		&$\times$\\ \hline
		\rule[16pt]{0pt}{0pt}
    $\sum m_i $&NH & $\geq 190$\,meV &$ \geq 165$\,meV&
    $\geq 195$\,meV&  $\geq140$\,meV& ---\\ 
		\rule[14pt]{0pt}{0pt} 
		at common $\tau$	& IH &$\geq 680$\,meV&---&$\geq 420$\,meV&
		 ---&---\\ \hline
		\rule[16pt]{0pt}{0pt}
		$\delta_{CP}^\ell$ 
		&NH & $0$--$360^\circ$ &$0$--$20^\circ$,  &  $110^\circ$--$125^\circ$, &$0$--$20^\circ$, $340^\circ$--$360^\circ$, & --- \\ 
		\rule[14pt]{0pt}{0pt} 
		at best fit of $\sin^2\theta_{23}$		& & & $340^\circ$-$360^\circ$ &
		$235^\circ$--$250^\circ$ & 	 $60^\circ$-$100^\circ$, $260$--$300^\circ$&   \\
		\rule[14pt]{0pt}{0pt} 
		in $\sum m_i\leq 120$\,meV& IH &---&---&---&---&---\\ \hline	
		\rule[16pt]{0pt}{0pt} 
		$\langle m_{ee}\rangle$	&NH & $1.5$--$28$\,meV & $16$--$31$\,meV&
		$15$--$30$\,meV & $22$--$30$\,meV& ---\\  
		\rule[14pt]{0pt}{0pt}
		in $\sum m_i\leq 120$\,meV	& IH & --- &---&$22$--$24$\,meV&---&---\\ \hline	 	 
	\end{tabular}
	\caption{Summary of characteristic predictions for NH and IH of cases I -- V.
	}
	\label{tb:weight6}
\end{table}

Finally, we  discuss briefly the case of IH of neutrino masses.
 Indeed, there is the common region of $\tau$ in  quarks and leptons for the case I of Eq.(\ref{ME642}), where
 the predicted $\sin^2\theta_{23}$ is larger than $0.6$, and
 $\delta_{CP}^\ell$ is around  $70^\circ$  and $290^\circ$.
 However,  this case  is completely excluded because
 the sum of neutrino masses is larger than  $680$\,meV.
   Observed three mixing angles of leptons are not reproduced
   below $\sum m_i\leq 600$\,meV.
    In the case I\hspace{-.01em}I of Eq.(\ref{ME622}), there is no common region
 of $\tau$ in quarks and leptons.
  Observed three mixing angles of leptons are not reproduced
   unless  the sum of neutrino masses is larger than  $200$\,meV.
 In the case I\hspace{-.15em}I\hspace{-.15em}I, 
 it is found the common region of $\tau$ in quarks and leptons,
  but the sum of neutrino masses is larger than  $420$\,meV.
  Observed three mixing angles of leptons are  reproduced
   below   $\sum m_i\leq 120$\,meV,
   where $\sin^2\theta_{23}\simeq 0.49$--$0.52$,
    $\delta_{CP}^\ell= 50^\circ$--$60^\circ$, $160^\circ$--$170^\circ$, $190^\circ$--$200^\circ$, $300^\circ$--$310^\circ$,
      and 
      $\langle m_{ee}\rangle=22$-$24$\,meV.
      For cases I\hspace{-.01em}V and V,
       there are no common regions of $\tau$  in quarks and leptons.
       Below $\sum m_i\leq 120$\,meV, 
        observed three mixing angles of leptons are not reproduced
      in our scan  regions of 
       $|{\rm Re} [\tau]|\leq 0.1 $  and ${\rm Im} [\tau]\leq 1.12$.
       In conclusion, IH of  neutrino masses is unfavorable in our model.
      We summarize  the results of IH in Table 7.


\section{Summary and discussions}

In this work, we have studied both quark and lepton mass matrices
in the $A_4$ modular symmetry towards the unification of quark and lepton
flavors.
If  flavors of quarks and leptons are originated from a same two-dimensional compact space, 
the quarks and leptons have same flavor symmetry and the common  modulus  $\tau$.

The viable quark mass matrices are composed of
 modular forms of weights $2$, $4$ and $6$.
  It is remarked that $\tau$ is close to $i$,
 which is a fixed point in the fundamental region of SL$(2,Z)$,
 and  the CP symmetry is not violated.
 Indeed,  we reproduce  the observed CP violation of the quark sector
 at $\tau$ which is deviated a little bit from  $\tau=i$.
  
  The  charged lepton mass matrix
  is also given by using modular forms of weights $2$, $4$ and $6$.
  In order to study the quantitative dependence of our result
  on weights of the right-handed charged leptons, we have examined 
   five cases I--V of them.
  The neutrino mass matrix is generated  in terms of 
   the modular forms of weight $4$  through the Weinberg operator.  
  
  Our lepton mass matrices are also  consistent with the observed
  mixing angles at  $\tau$ close to $i$ for NH of neutrino masses.
 It is found that  
   allowed regions of $\tau$ of quarks and  leptons  overlap each other
   for all cases of the charged lepton mass matrix.
    The sum of neutrino masses is  crucial to test
    the common $\tau$ for quarks and leptons.
   The predicted  minimal sum of neutrino masses $\sum m_i$
   is  $140$\,meV at the common $\tau$ in the case  I\hspace{-.01em}V.
   If the  cosmological upper-bound of the sum of neutrino masses, 
   $120$\,meV will be confirmed,  the common region of $\tau$ of  quarks and leptons vanishes.
    However, the allowed region of $\tau$ in both quark and lepton sectors
     could be   shifted to a certain extent by some corrections such as 
     the SUSY breaking effect through
     threshold corrections to  masses and mixing angles.
    The appreciable shift of  $\tau$ could be also  occurred by
    the modification of the quark and lepton mass matrices.
    We need further investigation of the  mass matrices to reproduce the observed  CKM and PMNS  on the common $\tau$. 
    
As well known, the modulus $\tau=i$ is a fixed point, which is invariant
 in the  $\mathbb{Z}_2^{S}=\{ I, S \}$ group.
In our numerical results, the modulus $\tau$ is fixed close to $\tau=i$,
which suggests the approximate residual symmetry  
in the quark and lepton mass matrices. 
Some corrections could violate the exact symmetry.
The group theoretical investigation will be presented in the near future.
It is also emphasized that  
the spontaneous CP violation in Type IIB string theory  is possibly realized nearby $\tau=i$, where the moduli
stabilization as well as the calculation of Yukawa couplings is performed in a controlled way \cite{Kobayashi:2020uaj}.
Thus, our phenomenological result of $\tau$ may be favored 
 in the theoretical investigation.

 It may be useful to note  that  IH of  neutrino masses
 is unfavorable in our framework.
 Our study provides a phenomenological new aspect towards the unification of the quark and lepton flavors in terms of  the modulus $\tau$.

\section*{Acknowledgments}
This research was supported by an appointment to the JRG Program at the APCTP through the Science and Technology Promotion Fund and Lottery Fund of the Korean Government. This was also supported by the Korean Local Governments - Gyeongsangbuk-do Province and Pohang City (H.O.). 
H. O. is sincerely grateful for the KIAS member. 


\appendix
\section*{Appendix}

\section{Tensor product of  $A_4$ group}
We take the generators of $A_4$ group as follows:
\begin{align}
\begin{aligned}
S=\frac{1}{3}
\begin{pmatrix}
-1 & 2 & 2 \\
2 &-1 & 2 \\
2 & 2 &-1
\end{pmatrix},
\end{aligned}
\qquad 
\begin{aligned}
T=
\begin{pmatrix}
1 & 0& 0 \\
0 &\omega& 0 \\
0 & 0 & \omega^2
\end{pmatrix}, 
\end{aligned}
\end{align}
where $\omega=e^{i\frac{2}{3}\pi}$ for a triplet.
In this base,
the multiplication rule of the $A_4$ triplet is
\begin{align}
\begin{pmatrix}
a_1\\
a_2\\
a_3
\end{pmatrix}_{\bf 3}
\otimes 
\begin{pmatrix}
b_1\\
b_2\\
b_3
\end{pmatrix}_{\bf 3}
&=\left (a_1b_1+a_2b_3+a_3b_2\right )_{\bf 1} 
\oplus \left (a_3b_3+a_1b_2+a_2b_1\right )_{{\bf 1}'} \nonumber \\
& \oplus \left (a_2b_2+a_1b_3+a_3b_1\right )_{{\bf 1}''} \nonumber \\
&\oplus \frac13
\begin{pmatrix}
2a_1b_1-a_2b_3-a_3b_2 \\
2a_3b_3-a_1b_2-a_2b_1 \\
2a_2b_2-a_1b_3-a_3b_1
\end{pmatrix}_{{\bf 3}}
\oplus \frac12
\begin{pmatrix}
a_2b_3-a_3b_2 \\
a_1b_2-a_2b_1 \\
a_3b_1-a_1b_3
\end{pmatrix}_{{\bf 3}\  } \ , \nonumber \\
\nonumber \\
{\bf 1} \otimes {\bf 1} = {\bf 1} \ , \qquad &
{\bf 1'} \otimes {\bf 1'} = {\bf 1''} \ , \qquad
{\bf 1''} \otimes {\bf 1''} = {\bf 1'} \ , \qquad
{\bf 1'} \otimes {\bf 1''} = {\bf 1} \  .
\end{align}

More details are shown in the review~\cite{Ishimori:2010au,Ishimori:2012zz}.


\section{Majorana and Dirac phases and $\langle m_{ee}	\rangle $
in  $0\nu\beta\beta$ decay }

Supposing neutrinos to be Majorana particles, 
the PMNS matrix $U_{\text{PMNS}}$~\cite{Maki:1962mu,Pontecorvo:1967fh} 
is parametrized in terms of the three mixing angles $\theta _{ij}$ $(i,j=1,2,3;~i<j)$,
one CP violating Dirac phase $\delta _\text{CP}$ and two Majorana phases 
$\alpha_{21}$, $\alpha_{31}$  as follows:
\begin{align}
U_\text{PMNS} =
\begin{pmatrix}
c_{12} c_{13} & s_{12} c_{13} & s_{13}e^{-i\delta^\ell_\text{CP}} \\
-s_{12} c_{23} - c_{12} s_{23} s_{13}e^{i\delta^\ell_\text{CP}} &
c_{12} c_{23} - s_{12} s_{23} s_{13}e^{i\delta^\ell_\text{CP}} & s_{23} c_{13} \\
s_{12} s_{23} - c_{12} c_{23} s_{13}e^{i\delta^\ell_\text{CP}} &
-c_{12} s_{23} - s_{12} c_{23} s_{13}e^{i\delta^\ell_\text{CP}} & c_{23} c_{13}
\end{pmatrix}
\begin{pmatrix}
1&0 &0 \\
0 & e^{i\frac{\alpha_{21}}{2}} & 0 \\
0 & 0 & e^{i\frac{\alpha_{31}}{2}}
\end{pmatrix},
\label{UPMNS}
\end{align}
where $c_{ij}$ and $s_{ij}$ denote $\cos\theta_{ij}$ and $\sin\theta_{ij}$, respectively.

The rephasing invariant CP violating measure of leptons \cite{Jarlskog:1985ht,Krastev:1988yu}
is defined by the PMNS matrix elements $U_{\alpha i}$. 
It is written in terms of the mixing angles and the CP violating phase as:
\begin{equation}
J_{CP}=\text{Im}\left [U_{e1}U_{\mu 2}U_{e2}^\ast U_{\mu 1}^\ast \right ]
=s_{23}c_{23}s_{12}c_{12}s_{13}c_{13}^2\sin \delta^\ell_\text{CP}~ ,
\label{Jcp}
\end{equation}
where $U_{\alpha i}$ denotes the each component of the PMNS matrix.

There are also other invariants $I_1$ and $I_2$ associated with Majorana phases
\begin{equation}
I_1=\text{Im}\left [U_{e1}^\ast U_{e2} \right ]
=c_{12}s_{12}c_{13}^2\sin \left (\frac{\alpha_{21}}{2}\right )~, \quad
I_2=\text{Im}\left [U_{e1}^\ast U_{e3} \right ]
=c_{12}s_{13}c_{13}\sin \left (\frac{\alpha_{31}}{2}-\delta^\ell_\text{CP}\right )~.
\label{Jcp}
\end{equation}
We can calculate $\delta^\ell_\text{CP}$, $\alpha_{21}$ and $\alpha_{31}$ with these relations by taking account of 
\begin{eqnarray}
&&\cos\delta^\ell_{CP}=\frac{|U_{\tau 1}|^2-
	s_{12}^2 s_{23}^2 -c_{12}^2c_{23}^2s_{13}^2}
{2 c_{12}s_{12}c_{23}s_{23}s_{13}}~ , \nonumber \\
&&\text{Re}\left [U_{e1}^\ast U_{e2} \right ]
=c_{12}s_{12}c_{13}^2\cos \left (\frac{\alpha_{21}}{2}\right )~, \qquad
\text{Re}\left [U_{e1}^\ast U_{e3} \right ]
=c_{12}s_{13}c_{13}\cos\left(\frac{\alpha_{31}}{2}-\delta^\ell_\text{CP}\right )~.
\end{eqnarray}
In terms of these parametrization, the effective mass for the $0\nu\beta\beta$ decay is given as follows:
\begin{align}
\langle m_{ee}	\rangle=\left| m_1 c_{12}^2 c_{13}^2+ m_2s_{12}^2 c_{13}^2 e^{i\alpha_{21}}+
 m_3 s_{13}^2 e^{i(\alpha_{31}-2\delta^\ell_{CP})}\right|  \ .
\end{align}


\end{document}